\documentstyle[12pt]{article}
\begin{document}

\title{Noether Currents for Bosonic 
Branes}

\author{
G. Arreaga${}^{(1)}$, R. Capovilla${}^{(1)}$
and J. Guven${}^{(2)}$ \vspace{1cm}\\
${}^{(1)}${\it Departamento de F\'{\i}sica}\\
{\it Centro de Investigaci\'on y de Estudios Avanzados del IPN}\\
{\it Apdo. Postal 14-740,07000 M\'exico, DF, MEXICO}
\vspace{.5cm}\\
${}^{(2)}${\it Instituto de Ciencias Nucleares}\\
{\it Universidad Nacional Aut\'onoma de M\'exico}\\
{\it Apdo. Postal 70-543, 04510 M\'exico, DF, MEXICO}}

\maketitle
\begin{abstract}
We consider a relativistic brane propagating in 
Minkowski spacetime described by any action which is local in 
its worldvolume geometry. We examine the conservation laws 
associated with the Poincar\'e symmetry of the background 
from a worldvolume geometrical point of the view. 
These laws are exploited to explore the structure of the
equations of motion. General expressions are provided   
for both the linear and angular momentum 
for any action depending on the worldvolume extrinsic 
curvature. The conservation laws are 
examined in perturbation theory.
It is shown how non-trivial solutions with vanishing 
energy-momentum can be constructed in higher order theories.
Finally, subtleties associated with boundary 
terms are examined in the context of the brane Einstein-Hilbert 
action.
\end{abstract}

\newpage

\section{Introduction} 

A brane is a relativistic extended object
propagating in a background spacetime, usually treated as fixed, of
some given dimension $N$. The spacetime trajectory of the brane is a 
timelike worldvolume of some lower dimension $D$:
for a relativistic string $D=2$; for a 
domain wall, or membrane, $D=3$, and so on.
The dynamics of the brane is described
by some local action constructed using scalars 
that characterize the geometry
of the worldvolume. The possibilities are limited 
by  worldvolume reparameterization invariance, 
ambient spacetime diffeomorphism invariance,
and, when $D<N-1$, invariance under rotations in the 
$N-D$ dimensional plane normal to the worldvolume.
The simplest action of this kind is the Dirac-Nambu-Goto 
[DNG] action, proportional to the volume swept out by the
brane in the course of its evolution.
This action depends only on first order
derivatives of the field variables, the
embedding functions that define the worldvolume.
Originally proposed by Dirac in the context of
an extensible model for the electron ($D=3$) \cite{Dirac,CHT}, 
it was later exploited by Nambu and Goto for 
the case of a relativistic
string ($D=2$) to  model  hadronic
matter \cite{NG}. It constitutes the point 
of departure in the construction 
of modern string theory \cite{GSW}. Higher dimensional  ($D>1$) DNG 
branes also play an increasingly important role as solitonic solutions of
this theory (see {\it e.g.} \cite{Bachas,Nicolai}). 

Despite 
significant developments in string theory, it would appear fair to claim 
that a comprehensive understanding of the 
dynamics of relativistic extended objects is still lacking.
In particular, there are many contexts where the DNG action is 
clearly inadequate, and it has been necesssary to consider
a more general action that includes higher-order geometrical scalars 
which depend on the curvature of the worldvolume; possibly even its 
derivatives. These additions do modify the dynamics substantially. 
The lowest order correction
to the dynamics of a purely geometrical object is described by an 
action quadratic in the extrinsic curvature.
For example, in a phenomenological approach,
the addition of such a `rigidity' term to the DNG action
was proposed in the eighties by Polyakov, and independently
by Kleinert,  as an improved
effective action for QCD  \cite{rs}. Earlier still, 
hamiltonians of this form
were studied in condensed matter physics 
as models to describe the 
mechanical properties of lipid membranes \cite{CH}. 
Higher-order corrections
can also arise in systematic approximations. For example,
in the context of the physics of topological
defects, curvature terms are induced by
considering an expansion in the thickness of
the defect \cite{finitewidth}.
The addition of such terms typically 
associates an energy penalty with the formation of a spike.
These corrections thus become important  
on short distance scales serving to 
smooth out the curvature singularities 
which will arise in the course of the DNG  dynamics. 
In Ref. \cite{bcrg}, an effective action of this form was
obtained explicitly for domain walls
by integrating out the microscopic degrees of freedom
of the underlying field theory which vary rapidly on the 
length scale of the wall. In the case of strings it has been argued
that  the lowest order effective action 
which describes the behavior in the neighborhood of 
a cusp must contain terms which are of  
fourth order in the extrinsic curvature 
of the worldvolume \cite{abgs}.
 
Another, different, source of higher-order terms
comes from supersymmetric branes.
They arise once an effective action is
obtained by integrating out the fermionic 
degrees of freedom 
 \cite{Polybook,Ambjorn,Viswa}.

In this paper, we explore the structure of the 
conservation laws associated
with the Poincar\'e invariance of a 
local brane action propagating in Minkowski 
spacetime. These occur as a consequence of the 
induced internal symmetries on the field theory described by the 
embedding of the brane worldvolume in spacetime. 

At one level, the determination of the 
associated conserved quantities 
is straightforward. One can write 
the action explicitly in terms of the 
embedding functions, perhaps in terms 
of a natural parametrization, if available.
Noether's theorem  then generates the 
conserved momentum and angular momentum. 
This procedure is perfectly adequate 
for a DNG object where the
lagrangian is a function of the intrinsic
geometry alone (see, for example, \cite{Scherk} and  \cite{GSW} for 
a DNG string, or \cite{Let} for higher dimensional DNG objects.)
If, however, the geometrical action involves either 
the intrinsic or the extrinsic curvature, so that the action 
depends on second derivatives
of the embedding functions
or higher, the manifest covariance of the geometrical action
we start out with gets mutilated in 
this straighforward approach, and the conserved 
quantities do not possess any obvious geometrical form. 

There are several ways to remedy this shortcoming. One way 
exploits the elegant formalism developed by Iyer and
Wald for diffeomorphism invariant theories 
\cite{IW}. This involves grouping higher derivative terms
into totally symmetric combinations, and
obtaining the general expressions for the Noether
charges.
For the case of branes, however, this procedure
does not result in the natural geometric quantities
of the worldvolume, and the translation is ackward.
An alternative strategy is to perform a hamiltonian
analysis. This is certainly necessary 
if one is interested in the canonical quantization
of the theory. However, 
since the extrinsic curvature of a brane can be 
considered, roughly, as a generalization 
to higher dimensional objects of the
acceleration of a relativistic particle,
one is dealing with the complicated hamiltonian analysis
of a higher derivative theory.
Another approach which is spacetime manifestly covariant  
was pioneered by Carter (see {\it e.g.} \cite{CarterMex}
and references therein). It focuses on 
the derivation of the stress-energy tensor for the theory, 
and the
conserved quantities are constructed by contraction
with the appropriate background Killing vector fields.
This approach is certainly convenient when external 
fields are present, but, in our opinion,
 it does not take  full
advantage of the natural geometric structures on the
worldvolume. 

In this paper we develop an independent approach, 
close in spirit to the one developed by Carter, 
but differing in our emphasis on the worldvolume geometry.
We exploit the geometrical formalism
introduced by two of the authors in \cite{Defo}, 
tailored to the worldvolume of the extended object,
to describe deformations of the worldvolume to express the 
variations in the worldvolume geometry  
induced by a Poincar\'e transformation in a  covariant way.
This will permit us to write down the 
corresponding conservation laws in a geometrical form
adapted to the worldvolume. 

Following Carter, we express the equations of motion in terms of the conservation of the linear momentum. However, we go one step further
by separating these equations into two sets,
obtained by projecting the local form of
the conservation law onto and
normal to the worldvolume. In doing this, we 
necessarily dismantle the worldvolume 
divergence appearing in the conservation law,
which one might consider a step in the wrong direction.
The advantage, however, is that
the $N-D$ normal projections encode completely the worldvolume
diffeomorphism  invariant content of the former equations.
The associated gauge redundancy in the conservation laws is 
captured in the $D$ tangential projections or Bianchi identities.
When we cast the conservation law this way, 
the distinct dynamical roles 
played by the tangential and normal projections of
the linear momentum density in the theory also become 
explicit. It is then an easy matter 
to identify characteristics of the projections associated with 
a specific theory, such as, for example, the conditions
under which a theory  is conformally invariant.

We demonstrate how this structure can be exploited 
in perturbation theory.
In particular, we exploit the linearization of the conservation law
to provide a novel derivation of the linearized equations of motion, 
which, in addition to its 
technical merits, also throws light on the underlying structure.

We also examine angular momentum conservation. The structure of the spin
for a DNG action has a very special form. We demonstrate explicitly 
how new structures enter when actions of higher order are considered.
We point out that the Regge inequality for string theory
continues to hold along extremal solutions of certain higher 
order theories.

An important issue in the path integral formulation of 
the quantum theory is the identification of 
appropriate boundary conditions to be
imposed on the initial and final configurations in the variational
principle. These boundary conditions are largely a matter of choice
in the classical theory. An example is 
provided by the brane Einstein-Hilbert action
\cite{RT,DPR} which involves terms linear in second derivatives
of the embedding functions, so that the equations of motion,
in common with DNG theory, is also second order in derivatives. 
A spurious surface term occurs in the variational 
principle which signals problems in the
path integral quantization of the theory. 
We provide a geometric approach to analysing this problem
which is guided by the analogous problem in General Relativity
\cite{York,Hawking}. 

It should be mentioned that 
various special cases the Noether currents
of particular branes have been analyzed
in the literature, using different approaches.
For example, Letelier has considered a DNG membrane
($D=2$) propagating in four-dimensional Minkowski
spacetime, with the addition of a term porportional
to the worldvolume intrinsic curvature, the 
brane Einstein-Hilbert action \cite{Letelier}.
A relativistic string with arbitrary curvature
corrections has been studied by Boisseau and 
Letelier \cite{BL}. The more general case of a brane action
at most quadratic in the extrinsic curvature
in an arbitrary background spacetime 
is the subject of a detailed analysis by Carter in 
\cite{CarterCQG}. For an alternative treatment of extrinsic curvature
actions which uses the language
of differential forms, see Hartley and Tucker in 
\cite{HT}. We find complete agreement with their
treatment, once details of notation have been 
taken into account.

As is well known, the expressions for the momenta themselves
are ambigous. One is always free to add a total divergence
to the lagrangian, which contributes to the momenta
without affecting the dynamics. Moreover, one
can add a term that is identically conserved to the
momentum densities. We will keep in mind this freedom,
and we will have occasion to use it for example in the  specific
case of the brane Einstein-Hilbert action, 
but we will not develop the general
formalism to treat it in full 
generality. For the interested reader, this subject
and its consequences on the global structure of
the space of solutions of a relativistic
theory is developed
in detail {\it e.g.} by  Anderson and Torre in \cite{Anderson,AT,Torre}.

This paper is organized as follows. In the next section,
we define the linear
and angular momentum for a general brane.
In Sect. 3, we show how the equations of motion
can be expressed in terms of the conservation of the
linear momentum. The familiar case of a DNG
brane is the subject of Sect. 4, which serves
as a useful illustration of the formalism. 
In Sect. 5, we move on to the general
case of a rigid lagrangian depending on the extrinsic
curvature of the worldvolume. We obtain both the linear
and angular momentum for this class of theories.
In order to make our treatment more concrete
we specialize to low order specific examples 
in Sect. 6, including the Einstein-Hilbert action
for a brane. 
In Sect. 7, we provide  a novel approach to
the linearized equations of motion, in terms of the linearization
of the linear momentum density. The introduction of a surface term of the
York type in the brane Einstein-Hilbert action is the subject of Sect. 8.
We conclude in Sect. 9 with a brief discussion.
Finally, in an appendix, we sketch the extension of the formalism 
to lagrangians that depend on derivatives of the extrinsic curvature.

\section{Conservation laws}

Let us consider a brane, 
of dimension $D-1$, propagating in a
fixed background Minkowski
spacetime of dimension $N$.
For simplicity, we suppose that this object is either 
infinite in extent, or closed. 
  In the case of an infinite
object, we will assume in the following that appropriate
fall-off conditions are chosen on the fields, so that the
formal expressions we derive actually exist.
The  extension of the 
formalism to objects with finite (timelike) boundaries,
or with loaded  edges, relevant for the treatment of
hybrid branes, such as a string with monopoles
at its edges, or a wall bounded by strings (see
{\it e.g.}  \cite{edges}), is 
straightforward, and will not be considered
in this paper.

The domain of integration  is the 
timelike worldvolume $m$,
given by the embedding functions 
\begin{equation}
x^\mu = Y^\mu(\xi^a)\,,
\label{eq:embe}
\end{equation}
where $x^\mu$ are local  coordinates
in the ambient Minkowski spacetime,
and $\xi^a$ local coordinates
for the worldvolume
($\mu,\nu, \cdots =
0, 1, \cdots N-1$, and 
$a,b, \cdots = 0, 1 \cdots, D-1$). 
The worldvolume is
described by the evolution of a $D-1$-dimensional
 brane  between 
fixed initial and final  (spacelike) configurations. 

The initial and final configurations
 are described completely by two 
spacelike hypersurfaces on the
worldvolume $m$, $\Sigma_{(i)}$ and $\Sigma_{(f)}$.  For
definiteness, we parametrize these hypersurfaces
by the embedding functions,
\begin{eqnarray}
\xi^a &=& X^a_{(i)}(u^A)\,, 
\label{eq:end1}
\\ 
\xi^a &=& X^a_{(f)} (u^A)\,,
\label{eq:end2}
\end{eqnarray}
respectively, where $u^A$ are local coordinates
for the spacelike hypersurfaces ($A,B, \cdots = 1, \cdots, D-1$).

It turns out that it is also convenient
to consider them 
as embedded directly in Minkowski
spacetime, via map composition, so that
{\it e.g.} with  (\ref{eq:embe}) and
(\ref{eq:end1}) we have
\begin{equation}
x^\mu = X_{(i)}^\mu ( u^A ) = e^\mu{}_a X_{(i)}^a (u^A )\,,
\label{eq:end3}
\end{equation}
where we denote by 
\begin{equation}
e^\mu{}_a = {\partial Y^\mu \over \partial
\xi^a }\,,
\label{eq:tange}
\end{equation}
the $D$ tangent vectors to the worldvolume
$m$. We will also use 
\begin{equation}
\epsilon^\mu{}_A = {\partial X^\mu \over \partial u^A }\,,
\end{equation}
to denote the $D-1$ tangent vectors to $\Sigma_{(i)}$ or $\Sigma_{(f)}$ 
as embedded in spacetime. 

The union of these two hypersurfaces is the boundary 
$\partial m$ of the worldvolume $m$, with the 
understanding that the natural orientation 
of the initial hypersurface $\Sigma_{(i)}$
is opposite to that of $\Sigma_{(f)}$.
We consider only oriented branes.

We now make an infinitesimal deformation of the 
embedding functions for the worldvolume $m$,
\begin{equation}
Y^{\mu}(\xi) \to  Y^{\mu}(\xi) + \delta Y^{\mu}(\xi) \,.
\label{eq:def}
\end{equation}
This displacement can be seen as a diffeomorphism
of the ambient spacetime, and it
will induce a deformation of the worldvolume geometry.

We decompose an 
arbitrary infinitesimal deformation of the 
embedding $\delta Y^\mu$ into its parts 
tangential and normal to the worldvolume, 
\begin{equation}
\delta Y^{\mu}= \Phi^a e^{\mu}{}_a + \Phi^i n^{\mu}{}_i\,,
\label{eq:vd}
\end{equation}
where the tangent vectors  $ e^\mu{}_a$ 
have been defined earlier, by Eq. (\ref{eq:tange}), 
and $n^\mu{}_i$
are the $N-D$ vectors normal to the worldvolume
$m$ ($i,j,\cdots =
1,2, \cdots, N-D$). These are defined by
\[
\eta_{\mu\nu} e^\mu{}_a n^\nu{}_i = 0\,,
\]
where $\eta_{\mu\nu}$ is the Minkowski metric, with
only one minus sign, which will be used to raise
and lower spacetime indices.
We choose to normalize the normal vector fields 
as
\[
\eta_{\mu\nu} n^\mu{}_i n^\nu{}_j =  \delta_{ij}\,,
\]
where $\delta_{ij}$ is the Kronecker delta.

The worldvolume geometry is described completely by 
the induced metric $\gamma_{ab}$, 
the extrinsic curvature $K_{ab}{}^i$, 
 and the extrinsic twist
$\omega_a{}^{ij}$, when the appropriate
integrability conditions are satisfied. 
The induced metric is defined by
\begin{equation}
 \gamma_{ab} = e^\mu{}_a
e^\nu{}_b \eta_{\mu\nu}\,.
\end{equation}
This metric, together with its inverse $\gamma^{ab}$,
will be used to lower and raise worldvolume indices.
The quantity $K_{ab}{}^i = K_{ba}{}^i $ is the 
extrinsic curvature along the $i$-th normal
vector field $n^\mu{}_i$,
\begin{equation}
K_{ab}{}^i = - n_{\mu}{}^i \partial_a e^\mu{}_b\,.
\end{equation}
The extrinsic twist $\omega_a{}^{ij}$, defined by
\begin{equation}
\omega_a{}^{ij} = \eta_{\mu\nu} n^{\mu\,j} \partial_a
n^{\nu\,i}\,,
\end{equation} 
 is the
connection associated with covariance 
under normal rotations. 

To evaluate the variation of these quantities under
a worldvolume deformation,
we will exploit the covariant formalism describing
deformations of the worldvolume geometry developed in Ref.\cite{Defo}. 
In this approach, the effect of the deformation on geometrical tensors 
is covariant not only with respect to reparameterizations
 of the worldvolume, 
but also with respect 
to local rotations of the normals $n^{\mu\,i}$.

We consider a local action, depending on the
embedding functions $Y^\mu$, which is 
both invariant under worldvolume reparametrization, and 
under rotations of the normals,
\begin{equation}
S [Y ] = \int_m \sqrt{-\gamma} L  
\end{equation}
(For convenience of notation, we have absorbed the 
worldvolume differential
$d^{D} \xi$ into the integral sign. We will do likewise
for the integrals over the boundary $\partial m$. )
The lagrangian  $L$  is constructed locally
from the geometry of the worldvolume $m$
\begin{equation}
L = L (\gamma_{ab}, K_{ab}{}^i , 
\widetilde\nabla_a K_{bc}{}^i, \cdots )\,,
\end{equation}
where we denote by $\widetilde\nabla_a$
the covariant derivative under rotation of the
normal vector fields introduced in \cite{Guven}. 
For an arbitrary normal vector 
$\Psi^i$, it is defined by 
\begin{equation}
\widetilde\nabla_a \Psi^i = \nabla_a \Psi^i - \omega_a{}^i{}_j
\Psi^j\,,
\end{equation}
where $\nabla_a$ denotes the (torsionless)
worldvolume covariant derivative
compatible with $\gamma_{ab}$.

The infinitesimal variation of the action which is 
induced by a worldvolume deformation
can always be decomposed into its 
tangential and normal  parts, 
\begin{equation}
\delta S = \delta_{\parallel} S + \delta_{\perp} S\,.
\end{equation}

Away from the boundary, the tangential deformation 
can be identified with a
diffeomorphism of $m$, since $\delta_{\parallel}S $ is 
a boundary term. 
Let us examine explicitly
how this occurs. We note that $\delta_ {\parallel} f 
= \Phi^a \partial_a f$ for any scalar function 
$f(\xi)$ defined on the worldvolume, $m$. In addition,  
under a tangential deformation, the induced metric on $m$
tranforms as a Lie derivative, 
\begin{equation}
\delta_ {\parallel} \gamma_{ab} = \nabla_a \Phi_b + \nabla_b \Phi_a\,.
\end{equation}
Thus
\begin{equation}
\label{eq:tangdef}
\delta_ {\parallel}\sqrt{-\gamma} = \sqrt{-\gamma}\; \nabla_a\Phi^a\,.
\end{equation}
For an arbitrary variation of the action, we have 
\begin{equation}
\delta S= 
\int_{m}
\left\{ \left( \delta \sqrt{-\gamma} \right) 
\; L
+ \sqrt{-\gamma} 
\; \left( \delta L \right)  \right\}\,.
\label{eq:arbi}
\end{equation}
A tangential deformation of the 
worldvolume thus always results in a pure 
divergence,
\begin{equation}
\delta_{\parallel} S = \int_m \sqrt{-\gamma}
\, \nabla_a 
\left(  L \Phi^a \right)
=  \int_{\partial m} \sqrt{h}
\; L \; \eta_a \,
\Phi^a  \,,
\label{eq:tpda}
\end{equation}
where we have used Stokes theorem in the second
equality, $h$ is the determinant of the metric
$h_{AB}$ induced on  $\partial m$ by the embeddings described by
Eq. (\ref{eq:end3}), 
\begin{equation}
h_{AB} =  \epsilon^\mu{}_A \epsilon^\nu{}_B \eta_{\mu\nu}\,,
\end{equation}
and  $\eta^a$ is the 
unit timelike normal on $\partial m$ pointing into $m$,
{\it i.e.} the  N-velocity of an observer sitting on 
the hypersurface $\Sigma$.
A diffeomorphism of $m$ can only move its boundary.

The action is stationary with respect to 
tangential deformations of the worldvolume with 
a vanishing normal component, 
$\eta_a\Phi^a =0$ on $\Sigma_{(i)}$ and $\Sigma_{(f)}$.
The remaining components of $\Phi^a$ may range freely. 
These are precisely the components that 
generate a diffeomorphism of the spacelike 
configurations:
the initial and final configurations are fixed, but {\it not} the coordinates
chosen to describe these configurations.

Whereas the tangential variation of the action 
is simple, the normal variation is, in general,
 non-trivial.
The normal deformation can always be cast in the form
\begin{equation}
\delta_{\perp} S =\int_m \sqrt{-\gamma} \;
\left[ {\cal E}_i( L ) \; \Phi^i
+   
 \; \nabla_a \; \Pi^a{}_{i}[\Phi^i]  \right]
\,,
\label{eq:perpdef}
\end{equation}
{\it i.e.} as a worldvolume part, and a pure divergence.
Here ${\cal E}_i ( L)$ is the Euler-Lagrange derivative of 
$ L$ projected onto the normals $n^\mu{}_i$ to the
worldvolume;
$\Pi^a{}_{i}$ is a linear differential operator defined on $m$ which arises when the worldvolume gradients of the normal deformation in the worldvolume bulk are confined
to a pure divergence, using integration by parts. The argument of the operator
$\Pi^a{}_{i}$ is indicated within the square bracket. 
We can use  the divergence theorem in the second term,
to obtain,
\begin{equation}
\delta_{\perp} S =\int_m \sqrt{-\gamma} \;
{\cal E}_i( L ) \; \Phi^i
+  \int_{\partial m} \sqrt{h} 
 \; \eta_a \; \Pi^a{}_{i}[\Phi^i] 
\,.
\label{eq:perpdef1}
\end{equation}
When the classical equations of motion are satisfied,
\begin{equation}
{\cal E}_i (L)=0\,,
\end{equation}
 the action is stationary with respect to 
normal deformations of $m$. A well posed
variational problem requires the vanishing of the boundary
terms on the initial and final configurations.  
The variational principle restricted to normal 
deformations gives the classical dynamics.

So far, we have considered arbitrary deformations
of the embedding functions. 
Let us now specialize to an infinitesimal Poincar\'e transformation, 
\begin{equation}
\label{eq:pd}
\delta Y^{\mu} = \epsilon^{\mu} +\omega^{\mu}{}_{\nu} Y^{\nu}\,, 
\end{equation}
where $\epsilon^\mu$ is an infinitesimal
constant translation,  and 
an infinitesimal
Lorentz transformation 
is given by
$ \Lambda^\mu{}_\nu = \delta^\mu_\nu + \omega^\mu{}_\nu $ ,
with
$\omega_{\mu\nu}= -\omega_{\nu\mu}$.
We decompose $\delta Y^\mu$ according to Eq.(\ref{eq:vd}).
For an infinitesimal spacetime translation $
\delta Y^\mu = \epsilon^\mu$, we have
\[
\Phi_i = n^{\mu}{}_i  \epsilon_{\mu} \,,\quad
\Phi_a = e^{\mu}{}_a \epsilon_{\mu} \,.
\]
Substituting this into
Eqs.(\ref{eq:tpda}) and 
(\ref{eq:perpdef}), and summing, 
the variation of the action associated
with a spacetime translation can be cast 
in the form,
\begin{equation}
\label{eq:gfdat1}
\delta S = \epsilon_{\mu}
\int_m \left[ \sqrt{-\gamma} \; {\cal E}^i (L) \;  n^\mu{}_i
+  \nabla_a {\cal P}^{a\mu}
\right]
\,.
\end{equation} 
The worldvolume vector density of weight
one ${\cal P}^{a\,\mu}$ is given by 
\begin{equation}
{\cal P}^{a\,\mu} = \sqrt{-\gamma}
\left( \Pi^a{}_{i}[ n^{\mu\,i}] +  L e^{\mu\,a}\right)
\,.
\label{eq:pam}
\end{equation} 
This expression for the variation of the action
plays a central role in what follows.
While tangential deformations do not participate
in the variational derivation of the equations of motion, 
we see that they do contribute in an essential way to the construction 
of conserved quantities.

The total boundary  contribution associated with a 
translation, using the divergence
theorem, is 
\begin{equation}
\label{eq:gfdat}
\delta S = \epsilon_{\mu}\int_{\partial m}  \;
\eta_a \, {\cal P}^{a\,\mu}\,,
\end{equation} 
with the understanding that ${\cal P}^{a\mu}$ is to
be evaluated at the boundary. This integral is well
defined since $ \eta_a {\cal P}^{a\mu}$ is
a density
of weight one when evaluated there.

Similarly, for an infinitesimal Lorentz transformation,
 we have
\[
\Phi_i =  \omega_{ \mu \nu}  n^{\mu}{}_i Y^{\nu}\,,\quad
\Phi_a =  \omega_{ \mu \nu} e^{\mu}{}_a  Y^{\nu}\,.
\]
and in this case 
the variation of the action associated with a Lorentz
transformation reduces to
\begin{equation}
\delta S = \omega_{\mu \nu } 
\int_m \left[ \sqrt{-\gamma} \; {\cal E}^i (L)
\; n^\mu{}_i \; Y^\nu
+  \nabla_a {\cal M}^{a\,\mu\nu} \right]\,,
\label{eq:varang3}
\end{equation}
where the worldvolume vector
 density of weight one ${\cal M}^{a\,\mu\nu} $
is
given by
\begin{equation}
{\cal M}^{a\,\mu\nu} = 
{1\over 2} \sqrt{-\gamma}
\left[ \Pi^a{}_{i}[n^{\mu\,i} Y^{\nu}] + L\, e^{\mu\,a} 
Y^\nu
- (\mu\leftrightarrow \nu) \right]\,.
\end{equation}
The boundary contribution to $\delta S$ gives,
\begin{equation}
\label{eq:gfdar}
\delta S =
\omega_{\mu \nu} \int_{\partial m} \;
\eta_a {\cal M}^{a\,\mu\nu}\,.
\end{equation}

In our derivation of the expressions for ${\cal P}^{a\,\mu}$ and 
${\cal M}^{a\,\mu\nu}$ it is important to emphasize that
we did not enforce any boundary conditions on the induced variations.
Indeed, it would be an error to attempt to enforce the 
boundary conditions which are appropriate for the variational 
derivation of the equations of motion.

Let us now
suppose that the equations of motion ${\cal E}_i (L)
=0$ are satisfied. 
We have 
\begin{equation}
\label{eq:linang}
\delta S= \epsilon_{\mu} \left[
P^{\mu}(\Sigma_{(f)})- P^{\mu}(\Sigma_{(i)}) \right]
+ 
\omega_{\mu \nu} \left[
M^{\mu \nu}(\Sigma_{(f)}) -M^{\mu \nu}(\Sigma_{(i)})
\right]
\;.
\end{equation}
The linear momentum $P^{\mu} (\Sigma )$ 
of the spatial hypersurfaces is defined by
\begin{equation}
P^{\mu} (\Sigma) =  \int_{\Sigma} \; 
 \; \eta_a \, {\cal P}^{a\,\mu}\,.
\label{eq:genm}
\end{equation}
We identify therefore
${\cal P}^{a\,\mu}$ as the linear 
momentum density.
From Eq. (\ref{eq:linang}) on we depart from our 
earlier convention, understanding  
the unit normal  $\eta^a$ to be future pointing
when referring to a spacelike hypersurface on $m$.

We emphasize that $P^\mu (\Sigma )$ is a quantity
associated with the spacelike hypersurface
$\Sigma$.
In fact, it is possible to express it purely in
terms of the geometry of $\Sigma$ itself.
We refrain from showing it explicitly, since
the investment in additional formalism
is not compensated by a corresponding gain
in information. On the
other hand, the density ${\cal P}^{a\mu}$ 
leads a double life:  depending on circumstances, it lives
either on the worldvolume $m$, or on its boundary
$\partial m$, as exemplified by Eqs. (\ref{eq:gfdat1})
 and (\ref{eq:gfdat}). 

We define the angular momentum $M^{\mu\nu} (\Sigma )$ 
by
\begin{equation}
\label{eq:genam}
M^{\mu \nu}(\Sigma) = 
\; \int_\Sigma  \;
\eta_a \, {\cal M}^{a\,\mu\nu}\,,
\end{equation}
which identifies ${\cal M}^{a\,\mu\nu}$ as the
angular momentum density.

It is useful to express ${\cal M}^{a\,\mu\nu}$ 
in the alternative form  
\begin{equation}
{\cal M}^{a\,\mu\nu} =
\frac{1}{2} 
\left[ 
{\cal P}^{a \mu} X^{\nu} + \pi^a{}_{i} [ n^{\mu\,i} Y^{\nu}]
- ( \mu \leftrightarrow \nu )\right]\,,
\label{eq:angpi}
\end{equation} 
where we introduce
 \begin{equation}
\label{eq:pi2}
\pi^a{}_{i} [ n^{\mu\,i} Y^{\nu}] 
= \sqrt{-\gamma} \left( \Pi^a{}_{i}[n^{\mu\,i} Y^{\nu}] - 
n^{\mu\,i} \Pi^a{}_{i }[X^{\nu}] \right) \,.
\end{equation} 
The antisymmetric part of 
$\pi^a{}_{i} [ n^{i\,\mu} X^{\nu}]$ denotes that part of 
${\cal M}^{a\,\mu\nu}$ which is not determined completely by the linear momentum density, ${\cal P}^{a\,\mu}$.
As we show below, this is precisely the part that is 
interesting in higher derivative theories.

If the action is Poincar\'e invariant, so that 
$\delta S =0$,
we have
\[
P^{\mu}(\Sigma_{(f)})=  P^{\mu}(\Sigma_{(i)})
\]
and
\[
M^{\mu \nu}(\Sigma_{(f)}) = M^{\mu \nu}(\Sigma_{(i)}) \,.
\]
However, $\Sigma_{(i)}$ and $\Sigma_{(f)}$ are arbitrarily chosen initial and 
final configurations of the brane.
Thus 
$P^\mu(\Sigma)$ and $M^{\mu\nu}(\Sigma)$ are both independent of 
the spacelike hypersurface
$\Sigma$. In this extremely broad sense,
both $P^{\mu}$ and $M^{\mu \nu}$ are constants of the motion.

Our treatment so far has been entirely general. 
We have not indicated how to evaluate either
the Euler-Lagrange derivatives of 
$L$, or the differential operator, $\Pi^a{}_{i}$. 
We will consider some concrete examples,
beginning with the most elementary, 
in  Sect. 4.

\section{Equations of motion from momentum
conservation}

Before we consider explicit examples, in this
section we show how the equations of motion
can be expressed in terms of the 
conservation of the linear momentum.

From the variation of the
action under a spacetime translation, Eq. 
(\ref{eq:gfdat1}), when
the action is invariant under  translations,
so that
$\delta S = 0$ on the left hand side, 
 we have
 \begin{equation}
\sqrt{-\gamma} \; {\cal E}^i (L) \; n^\mu{}_i =
- \nabla_a {\cal P}^{a\mu}\,.
\label{eq:emdp}
\end{equation}
This says that $\nabla_a {\cal P}^{a\mu}$ is normal
to the worldvolume, and the equations of motion 
imply the  conservation of the linear momentum density,
\begin{equation}
\nabla_a {\cal P}^{a\mu} = 0\,,
\label{eq:dpamu}
\end{equation}
and vice versa.
That the equations of motions can be restated
in terms of the conservation of linear momenta 
should not come as a surprise. In fact, it
is a special
case of the fact that the 
equations of motion  can be
expressed in terms of the conservation of the
stress-energy tensor
(for a relativistic string, see {\it e.g.}
\cite{AraDes}). As we mentioned in the introduction,
this is the approach
adopted by Carter in his
treatment of brane dynamics (see {\it e.g.} Ref.
\cite{CarterMex}, and references therein.). 

However, the form (\ref{eq:dpamu}) is not the most useful 
expression of the 
conservation law, as it  involves  the mixed 
spacetime-worldvolume
density, ${\cal P}^{a\,\mu}$;
it does not isolate its non-trivial part. 
It is possible, however, to express the 
equations of motion in purely worldvolume terms. 
First, we decompose the spacetime vector density ${\cal P}^{a\,\mu}$  
into its tangential and normal parts,
\begin{equation}
\label{eq:gep}
{\cal P}^{a\,\mu} =  {\cal P}^{ab}  e^{\mu}{}_b + 
{\cal P}^{a\,i} 
n^{ \mu}{}_i \,.
\end{equation} 
Note that, in general, the worldvolume tensor field 
density ${\cal P}^{a b}$ will not be
symmetric in its indices. 

The worldvolume covariant divergence of ${\cal P}^{a \mu}$ gives 
\begin{equation}
\nabla_a {\cal P}^{a\mu} = \left( \nabla_a {\cal P}^{ab} 
+  K^b{}_{a i}  {\cal P}^{ai} \right) e^{\mu}{}_b +
\left( \widetilde{\nabla}_a {\cal P}^{ai} -  
K_{ab}{}^i {\cal P}^{ab}  \right) n^{\mu}{}_i\,,
\label{eq:derden}
\end{equation}
where we have made use of the Gauss-Weingarten 
equations for the worldvolume $m$, (see {\it e.g.} \cite{Defo}),
\begin{eqnarray}
\nabla_a e^\mu{}_b &=& - K_{ab}{}^i \; n^\mu{}_i\,,
\label{eq:gw1} \\
\widetilde\nabla_a n^\mu{}_i &=& K_{ab\,i} \; e^{\mu\,b}\,.
\label{eq:gw2}
\end{eqnarray}
The worldvolume projections of the
expression (\ref{eq:emdp}), using Eq. (\ref{eq:derden}),
are therefore given by 
 \begin{eqnarray}
\widetilde{\nabla}_a {\cal P}^{ai} -  K_{ab}{}^i  {\cal P}^{ab}  &=& -
\sqrt{-\gamma} \; {\cal E}^i (L) \,.
\label{eq:em2} \\
\nabla_a {\cal P}^{ab} +  K^{b}{}_{a\,i}  {\cal P}^{ai} &=& 0\,,
\label{eq:em1}
\end{eqnarray} 
The first equation expresses the Euler-Lagrange 
derivative in a divergence form, so that the 
equations of motion take the form
\begin{equation}
\widetilde{\nabla}_a {\cal P}^{ai} -  K_{ab}{}^i  {\cal P}^{ab} 
= 0\,.
\label{eq:eom3}
\end{equation}
This equation is like a Gauss law for a $O(N-D)$
Yang-Mills ``electric field" ${\cal P}^{ai}$,
with ``source" $K_{ab}{}^i  {\cal P}^{ab} $.
Note that only the symmetric part of ${\cal P}^{ab}$
enters the equations of motion.

The Bianchi identity (\ref{eq:em1}) is a 
non-obvious integrability 
condition required for the existence 
of solutions of (\ref{eq:em2}). Here, also the antisymmetric
part of ${\cal P}^{ab}$ contributes.

We will exploit these expressions 
to examine the different role
played by the tangential and normal parts
of the momentum density. Moreover, as shown
below in Sect. 7, they allow for a novel approach to
 the linearization of the equations 
of motion.

We mentioned earlier that ${\cal P}^{ab}$
need not be symmetric. However, the
conservation of angular momentum,
$\nabla_a {\cal M}^{a\,\mu\nu} = 0$,
requires the anti-symmetric part to
vanish.

To conclude this section, let us point out that, in the
variational principle, one usually
keeps volume terms; boundary
terms, which contribute to the momenta, are thrown
away. The approach via conservation of momenta to
the equations of motion points to a complementary 
strategy: keep only boundary terms, since they also are sufficient
to reconstruct the dynamics.

\section{Dirac-Nambu-Goto action}

In order to illustrate the general formalism 
developed in the previous sections, 
let us begin with the familiar case
of a DNG brane.
The DNG action for a relativistic
extended object is 
\begin{equation}
S_{(0)} = - \mu \int_m \sqrt{-\gamma}\,,
\end{equation} 
where the constant $\mu$ is the brane tension. 
This is the simplest action one can write down
for such an object.
It depends only on the intrinsic geometry 
of the worldvolume. 
 
The normal deformation of this action is given by
\begin{equation}
\delta_{\perp} S_{(0)} =- \mu \int_m \sqrt{-\gamma}
\; K_i \; \Phi^i\,, 
\end{equation}
where we use the familiar expression
relating the Lie derivative along the normals
of the  volume element of $m$ to its 
mean extrinsic curvature  $K^i = \gamma^{ab} K_{ab}{}^i$,
\begin{equation}
\delta_{\perp} \sqrt{-\gamma}
 = \sqrt{-\gamma}\; K^i\; \Phi_i\,.
\label{eq:sqk}
\end{equation}
The tangential deformation of this action is simply
given by Eq. (\ref{eq:tpda}), with $L = - \mu$.

In this geometrical language, the equations 
of motion
are given by the vanishing of the mean extrinsic 
curvature,
\begin{equation}
\label{eq:K0}
- \mu K^i =0\,.
\end{equation}
The worldvolumes that extremize the DNG action are
extremal timelike surfaces.
This is a system of $N-D$ second-order hyperbolic 
partial differential equations for the 
embedding functions, $Y^\mu(\xi)$.
The appropriate boundary conditions in the variational principle are
$\eta_a \Phi^a =0$ on $\partial m$. Both 
$\Phi^i$ and $\Phi^a_\parallel
=\Phi^a - \eta_b \Phi^b \eta^a $ are arbitrary. We only need to fix
the initial and final hypersurface geometries. Anything more is superfluous.

To bring the equations of motion into a more familiar form, 
using the Gauss-Weingarten equations (\ref{eq:gw1}), 
we have that
\begin{equation}
K^i = - n^i_{\mu} \Delta Y^\mu\,,
\label{eq:K}
\end{equation}
where $\Delta $ is the worldvolume d'Alembert operator.
The tangential projections
of $\Delta Y^\mu$ vanish identically.
We can now peel Eq.(\ref{eq:K}), 
and its tangential counterpart to recover
the familiar harmonicity condition, 
\begin{equation}
\mu \Delta Y^\mu =0\,.
\end{equation}
Note that in this model there is no surface term arising from 
the normal variation, so that we have that
the operator introduced in Eq. (\ref{eq:perpdef})  vanishes
identically, $\Pi^{a}{}_i [\Phi^i ]=0$.
This is a feature which is unique to the DNG action.

The invariance of the DNG action under Poincar\'e transformations gives the linear momentum density
\begin{equation}
{\cal P}^{a\mu} = -\mu \sqrt{-\gamma} e^{\mu\,a}\,,
\label{eq:lddng}
\end{equation}
so that the total momentum $P^\mu (\Sigma )$ is given by
\begin{equation}
\label{eq:nglm}
P^\mu (\Sigma) = - \mu \int_{\Sigma} \sqrt{h} \; \eta^\mu\,,
\end{equation}
where $\eta^\mu = \eta^a e^\mu{}_{a}$ is  the 
unit velocity vector at a given point on $\Sigma$.
The momentum density ${\cal P}^{a\mu}$ is not only
tangent to the worldsheet, it also lies parallel to 
the tangent vector, $e^\mu{}_a$. In this sense, extremal 
surfaces, like geodesics, are self-parallel.
 
The linear momentum of a DNG brane is defined 
directly in terms of initial data on the spacelike
hypersurface $\Sigma$, without explicit reference to the 
worldvolume that will be generated by these
initial data. This expression for the linear
momentum of a DNG brane generalizes the expression for
a free relativistic massive particle, $P^\mu = m U^\mu$,
with $m$ its mass, and $U^\mu$ its unit  velocity.

The worldvolume projections
of ${\cal P}^{a\mu}$ are, respectively,
\begin{eqnarray}
{\cal P}^{ab}  &=& - \mu \sqrt{-\gamma} \gamma^{ab}\,,
\\
{\cal P}^{ai}  &=& 0\,. 
\end{eqnarray}
The vanishing of the normal part was to be expected --- the 
DNG equations of motion are of second
order in the embedding functions. Moreover,
the tangential projection is explicitly symmetric.
Substituting these projections into  
the worldvolume projections of the
linear momentum conservation equation,
Eqs. (\ref{eq:em2}) and (\ref{eq:em1}), we find that the first 
reproduces the extremal dynamics, $- \mu K^i=0$,
whereas the second is satisfied identically. 
We also note that ${\cal P}^{ab}$ is scale invariant,
under $\gamma_{ab} \to \Omega^2 \gamma_{ab} $,
if and only if $D=2$, {\it i.e.} for a relativistic
string. This is a consequence of the scale
invariance of the DNG string, which becomes
manifest in the Polyakov formulation of the
theory \cite{GSW}.

The angular momentum density is simply
\begin{equation}
{\cal M}^{a\mu\nu}
= {\cal P}^{a[\mu} X^{\nu ]}\,.
\end{equation}
When the equations of motion hold, this 
is automatically conserved.
The total angular momentum 
$M^{\mu\nu} (\Sigma )$ is given by
\begin{equation}
\label{ngam}
M^{\mu\nu} (\Sigma ) ={\mu \over 2} \int_{\Sigma}  
\sqrt{h} \,\left\{
\eta^{\mu} X^\nu 
- (\mu\leftrightarrow \nu)\right\}\,.
\end{equation}

Having covered the familiar
 case of a DNG brane, in the next section,
we move on to a less trivial class of applications.

\section{Extrinsic curvature actions}

In this section, we consider the simplest actions 
involving extrinsic curvature, described by some
lagrangian $L = L (\gamma^{ab},K_{ab}{}^i)$. 
The more general case of a lagrangian that
depends on derivatives of the 
extrinsic curvature as well 
can be treated along the same lines, 
and we will consider it in an Appendix.
The possibilities
are limited by the requirement
that the lagrangian must transform as a scalar under 
diffeomorphisms of the worldvolume, and under
rotations of the normals. In general, such theories
will involve derivatives of the embedding
functions higher than first, since the
extrinsic curvature generalizes to 
a brane the acceleration of a point particle. 

For a hypersurface,  $D=N-1$, 
a lagrangian proportional to (odd powers of) 
$K$ is admissible. 
For arbitrary co-dimension, however, the
lowest order lagrangian  invariant under 
normal rotations  is quadratic in $K_{ab}{}^i$,
\begin{equation}
L = -\mu + 
\alpha_1 K^i K_i + \alpha_2 K_{ab}{}^i K^{ab}{}_i\,,
\end{equation}
where $\alpha_1 $ and $\alpha_2$ are constants
that measure the rigidity of the brane.

The worldvolume scalars 
$K^i K_i$ and  $K_{ab}{}^i K^{ab}{}_i$ are 
not independent. The completely contracted 
Gauss-Codazzi equations in a flat spacetime 
background 
relates their difference to the worldvolume scalar curvature ${\cal R}$, with
\begin{equation}
\label{eq:gcc}
{\cal R}= K^i K_i - K^{a b\,i}K_{a b \,i}\,.
\end{equation}
When $D=2$, for a (closed) relativistic string, 
the action 
constructed from ${\cal R}$ is a topological invariant,
as follows from the Gauss-Bonnet theorem, so that the 
actions determined by the two quadratics are locally equivalent.

As before, the tangential variation of the action is straightforward, see Eq. (\ref{eq:tpda}).
We exploit the chain rule to write down the normal 
variation of the
lagrangian in terms of the variations of its arguments. We then have
\begin{equation}
\label{eq:defk}
\delta_{\perp} S = \int_m \sqrt{-\gamma} \left\{ 
K^i \Phi_i L + L^{a b}{}_i \; \left( \tilde\delta_{\perp} 
K_{a b}{}^i \right) 
+ L_{a b}\;  \left( \delta_{\perp} \gamma^{a b} \right)  \right\} \,,
\end{equation}
where we have used Eq. (\ref{eq:sqk}) for the first
term, and
 we have defined,
\begin{eqnarray}
L^{ab}{}_i= {\partial L \over \partial K_{ab}{}^i} = L^{ba}{}_i\,, \\
L_{ab}= {\partial L \over  \partial \gamma^{ab}}
 = L_{ba}\,.
\label{eq:ddef}
\end{eqnarray}
We now exploit the results of Ref.\cite{Defo},
specialized to a flat background,
 to express 
the normal variation of the inverse
induced metric and the extrinsic curvature as,
\begin{eqnarray}
\delta_{\perp} \gamma^{a b} &=& -2 K^{a b}{}_i \Phi^i\,,
\label{eq:dgab1}
\\
\tilde\delta_{\perp} K_{a b}{}^i 
&=& -\widetilde{\nabla}_a \widetilde{\nabla}_b \Phi^i 
+ K_{a c}{}^i K^c{}_{ b\,j} \Phi^j \,.
\label{eq:dkab}
\end{eqnarray}
The deformation operator $\tilde\delta_\perp$ 
acting on $K_{ab}{}^i$ is constructed analogously to 
$\widetilde\nabla_a$, and it involves a deformation
connection $\gamma_{ij} = - \gamma_{ji}$, so that
when acting {\it e.g.} on a vector under normal
rotations, it is defined by
\begin{equation}
\tilde\delta_{\perp} A^i = \delta_\perp A^i 
- \gamma_i{}^j A_j\,.
\end{equation}
This refinement is necessary
to ensure covariance of the deformation under
normal rotations, but coinciding with
$\delta_\perp $  when 
considering the deformation of normal rotation 
scalars. 
(For more detail the reader may refer to \cite{Defo}.)

We remove the  hessian of  $\Phi^i$ appearing in 
the second term on the right hand side of Eq.(\ref{eq:defk}), when we insert Eq. (\ref{eq:dkab}). 
The result is
\begin{eqnarray}
\delta_{\perp} S  &=& \int_m \sqrt{-\gamma}
\left\{ K^i  L - 2 K^{ab}{}^i L_{a b} 
- \widetilde{\nabla}_a \widetilde{\nabla}_b L^{a b}{}_i 
- K_{ac\,i} K_b{}^{c\, j} L^{a b}{}_j \right\} \Phi^i
\nonumber \\
&+& \int_{m} \sqrt{-\gamma}\,
\nabla_a \left\{ 
- L^{a b}{}_i \widetilde{\nabla}_b \Phi^i + 
\left( \widetilde{\nabla}_b L^{a b}{}_i \right) \Phi^i \right\}\,.
\label{eq:varperpak}
\end{eqnarray}
We identify the Euler-Lagrange derivative as
 \begin{equation}
{\cal E}_i (L)
 = - \widetilde\nabla_a\widetilde\nabla_b L^{ab}{}_i + 
L^{ab}{}_j K_{ac}{}^j K^c{}_{b\,i}  + 
L K_i -  2 L_{ab}K^{ab}{}_i \; .
\label{eq:elec}
\end{equation}
Generically, the Euler-Lagrange equations ${\cal E}^i =0$ are of second order
in derivatives of $K_{ab}{}^i$, so they
are of fourth order in derivatives of 
the embedding functions $Y^\mu$. 

Appropriate boundary conditions in the variational principle are that $\Phi^i=0$, in order to cancel
the last term in Eq. (\ref{eq:varperpak}), and,
since this implies already the 
vanishing of the derivative 
of $\Phi^i$ along $\Sigma$, in order
to cancel the next to last term in
Eq. (\ref{eq:varperpak}), we need only to
require $\eta^a\widetilde\nabla_a\Phi^i=0$ on $\partial m$,
independently of $L^{ab}{}_i$.

We identify
the operator $\Pi^a{}_{i}$ 
introduced in Eq.(\ref{eq:perpdef})
as the covariant `wronskian':
\begin{equation}
\Pi^a{}_i [\Phi^i] =
- L^{a b}{}_i \widetilde{\nabla}_b \Phi^i + 
\left( \widetilde{\nabla}_b L^{a b}{}_i \right) \Phi^i \,.
\end{equation}
In particular, for a $\Phi^i$ which corresponds to a background translation,
\begin{equation}
\epsilon_\mu \Pi^a{}_{i} [n^{\mu\,i}] =
 \epsilon_\mu \left[ 
- L^{a b}{}_i  K_{b}{}^{c\,i} e^\mu{}_c + 
\left( \widetilde{\nabla}_b L^{a b}{}_i \right) n^{\mu\,i}
\right]\,.
\end{equation}
where we exploit the Gauss-Weingarten equation
(\ref{eq:gw2})
to simplify the first term. We thus have from
Eq. (\ref{eq:pam}), the 
general expression for the linear momentum density, 
\begin{equation}
{\cal P}^{a\mu} = \sqrt{-\gamma} 
\left[ ( L \gamma^{ab} - L^{ac}_{i } K^b{}_{ c\,i}) e^\mu{}_{b}
+ \left( \widetilde{\nabla}_b L^{ab}{}_{i} \right) n^{\mu\,i} \right]\,.
\end{equation}
Unlike the DNG case, in general, the momentum density
${\cal P}^{a\mu}$ now possesses a component normal to
the worldvolume. 

We can write the tangential
part in a different way. First we separate the
quantity $L^{ac}_{i } K^b{}_{ c\,i}$
in its symmetric and anti-symmetric parts. It is easy
to show that the symmetric
part is given by
\begin{equation}
 L^{ca}{}_i K^{b}{}_{c}{}^i + L^{cb}{}_i K^{a}{}_{c}{}^i 
= 2 L^{ab}\,.
\label{eq:iden}
\end{equation}
Then, we can rewrite the tangential part 
of ${\cal P}^{a\mu}$ in the form
\begin{equation}
{\cal P}^{ab} = \sqrt{-\gamma}  \left(
L \gamma^{ab} - L^{ab} \right) + {\cal Q}^{ab} \,,
\label{eq:lproj3}
\end{equation}
where we isolate the anti-symmetric part of
${\cal P}^{ab}$,
\begin{equation}
{\cal Q}^{ab} = \sqrt{-\gamma} \left( 
L^{ca}{}_i K^{b}{}_{ c\,i} -  L^{cb}{}_i K^{a}{}_{ c\,i} 
\right) = 0\,,
\end{equation}
which vanishes identically
for the geometrical actions we consider.

In order to check in a non-trivial case that
the equations of motion can be expressed in
the form (\ref{eq:eom3}), we can substitute  the normal and tangential projections 
into the  left hand side of Eq.(\ref{eq:em2}), we obtain
the equations of motion in the form
(\ref{eq:elec}). Moreover, we can check
that Eq.(\ref{eq:em1}), corresponding to a worldvolume
translation, is in fact an identity.

Inspection of the form (\ref{eq:em2}) 
of the equations of motion,
or directly of Eq. (\ref{eq:elec}), 
shows that the necessary
and sufficient condition for the equations of motion
to be of second order in derivatives of the embedding
functions is simply  that the normal component
of the linear momentum density vanishes,
${\cal P}^{ai} = 0$.

The alternative expression (\ref{eq:lproj3}) for the 
tangential part of the linear momentum density
 is also useful for 
recovering the result that invariance
of the action under scale transformations
implies 
\begin{equation}
 {\cal P}^{ab} \gamma_{ab} = 0\,.
\label{eq:trss}
\end{equation}
To see this, consider that, using Eq. (\ref{eq:arbi}), the 
variation of the
action under a change of the worldvolume metric is
simply
\begin{eqnarray}
\delta S &=& \int_m \sqrt{-\gamma} \left[ L \gamma^{ab} 
- L^{ab} \right] \delta \gamma_{ab}
\nonumber \\
&=& \int_m {\cal P}^{ab} \delta \gamma_{ab}\,,
\end{eqnarray}
The second line follows from the fact that
only the symmetric part of ${\cal P}^{ab}$
enters. This identifies the symmetric 
part of ${\cal P}^{ab}$ as the
worldvolume stress-energy tensor.
When the induced metric undergoes a scale transformation,
 scale invariance of the action, $\delta S = 0$, implies
the tracelessness condition (\ref{eq:trss}), and viceversa.

Let us now move on to the angular momentum.
The operator appearing in Eq.(\ref{eq:pi2}) is
\begin{equation}
\pi^a{}_{i} [n^{i\,\mu}X^\nu] =
- L^{a b}{}_i  n^{\mu\,i} e^\nu{}_b \,,
\end{equation}
so that the angular momentum density
is given by
\begin{equation}
{\cal M}^{a\,\mu\nu} = 
\frac{1}{2}\; 
\left[ {\cal P}^{a\,\mu} X^{\nu} 
- \sqrt{-\gamma} \; L^{ab}{}_i \; n^{\mu\,i} \, e^\nu{}_{b} 
- (\mu\leftrightarrow \nu)\right]\,.
\label{eq:Mab}
\end{equation}
which does not involve derivatives 
in $K_{ab}{}^i$ 
other than those already contained in ${\cal P}^{a\,\mu}$.
The second term may be thought of as an effect of the
finite width of the worldvolume, when we go beyond
the DNG approximation. 
Its appearance is necessary to ensure 
conservation of angular momentum. Neither term alone is 
conserved. To see this, let us assume that the equations
of motion hold or, equivalently, that the linear momentum is
conserved, $\nabla_a {\cal P}^{a\mu} = 0$.
The divergence of the angular momentum density is
\begin{eqnarray}
\nabla_a {\cal M}^{a\mu\nu} = {1 \over 2}
[ {\cal P}^{a \mu} e^\nu{}_a &-& \sqrt{-\gamma} \; (\widetilde\nabla_a L^{ab}{}_i )
n^{\mu\,i} e^\nu{}_b - 
\sqrt{-\gamma} \; L^{ab}{}_i K_a{}^{c\,i} 
e^\mu{}_c e^\nu{}_b \nonumber \\
&+& 
\sqrt{-\gamma} \;
L^{ab}{}_i K_{ab\,j} n^{\mu\,i} n^{\nu\,j}
- (\mu\leftrightarrow \nu) ]\,,
\end{eqnarray}
where we have used the Gauss-Weingarten equations
(\ref{eq:gw1}), (\ref{eq:gw2}). In this expression,
all of the possible spacetime bivectors appear. 
However, if we express ${\cal P}^{a\mu}$ in 
terms of its projections, we obtain
\begin{equation}
\nabla_a {\cal M}^{a\mu\nu} = {1 \over 2}
\left[ 2 {\cal Q}^{ab} e^\mu{}_a e^\nu{}_b 
- \sqrt{-\gamma} \; L^{ab}{}_i K_{ab\,j} n^{\mu\,i} n^{\nu\,j}
- (\mu \leftrightarrow \nu)) \right]\,.
\end{equation}
The two terms are independent, the first is proportional 
to a bivector parallel to the worlvolume, the
second to a bivector normal to it. 
Therefore the conditions necessary for
conservation of angular momentum are
\begin{eqnarray}
{\cal Q}^{ab} &=& 0\,, 
\label{eq:cond1}\\
L^{ab}{}_{i} K_{ab\,j} - L^{ab}{}_j K_{ab\,i} &=& 0 
\label{eq:cond2}\,.
\end{eqnarray} 
The first condition was to be expected
from  classical elasticity theory \cite{LL}, and, as
mentioned above, is satisfied identically. 
The second condition is a new ``thickness" effect associated with
co-dimension $N-D>1$, and it is also satisfied identically 
for the geometrical actions we consider.

We also note that at this order,
${\cal M}^{a\,\mu\nu}$ does not involve any term 
proportional to $n^{\mu i} n^{\nu j} - (\mu \leftrightarrow
\nu)$, which is 
permitted if $N-D>1$. Such a term will show up in 
higher order theories, as we show in Appendix A.

\section{Extrinsic curvature actions: examples}

In order to make our discussion more concrete,
in this section we consider some specific examples
of the class of theories treated in the previous
section.
We begin with the simple case of a hypersurface
action of the form 
\begin{equation}
S_{(1)} = \alpha_0 \int_m \sqrt{\gamma} \;  K\,.
\end{equation}
Since $L_{ab} = 
\alpha_0 K_{ab}$ and $L^{ab}{}_\perp = \alpha_0 \gamma^{ab}$ (we use the symbol $\perp$ to denote
the only normal direction), from Eq. (\ref{eq:elec}),
we find the Euler-Lagrange derivative in the form
\begin{equation}
\label{eq:elk1}
{\cal E}_\perp = \alpha_0 ( K^2 - K^{ab} K_{ab} )  
= \alpha_0 {\cal R}\,,
\end{equation}
where we have used the contracted Gauss-Codazzi
equation (\ref{eq:gcc}). 
This action  is extremized by worldvolumes with
vanishing scalar Ricci curvature. It is a topological invariant (the 
winding number) for a pointlike trajectory.
The  equations
of motion, like the DNG case, are of second order in time derivatives
of the embedding functions. Indeed, 
the total linear momentum is given by
\begin{equation}
P^\mu (\Sigma )= \alpha_0 \int_{\Sigma} \sqrt{h}
 \; \eta_a \; 
\left( K \gamma^{ab} - K^{ab}\right) e^\mu{}_{b}\,,
\end{equation}
and we see that the linear momentum 
density is purely tangential, and manifestly
symmetric.
Moreover, reading off the quantity 
${\cal P}^{ab}$, and substituting in the equations
of motion in the form (\ref{eq:em1}), reproduces
${\cal E}_\perp = 0$, with the Euler-Lagrange
derivative given by Eq. (\ref{eq:elk1}).

Although the addition of this action to the DNG
action does not change the order of the equations
of motion, it {\it does} change the boundary conditions.
Now we need to require $\eta^a \widetilde\nabla_a \Phi^i
= 0$. We will discuss this issue in more detail
below, in the special case of a Einstein-Hilbert
brane action.

The total angular momentum is
\begin{equation}
M^{\mu\nu} (\Sigma )  = {1 \over 2} \int_{\Sigma} 
 \; \eta_a \; \left[ {\cal P}^{a\,\mu} X^{\nu} 
- \alpha_0 \sqrt{h} \; n^{\mu} \, e^{\nu\,a } 
- (\mu\leftrightarrow \nu) \right]\,.
\end{equation}
The second term is
proportional to the normal bivector
$ n^{\mu} \eta^{\nu} - n^\nu \eta^\mu $.
In the equation for angular momentum conservation,
$\nabla_a {\cal M}^{a\mu\nu} = 0$,
both terms are conserved separately. The first because
of the vanishing of ${\cal P}^{ai}$ and the symmetry of 
${\cal P}^{ab}$. The second is conserved automatically,
because the condition (\ref{eq:cond2}) is vacuus.
Alternatively, this can be checked directly using
the Gauss-Weingarten equations (\ref{eq:gw1}), 
(\ref{eq:gw2}), specialized to the
case of a hypersurface.

Let us now examine the case of an action quadratic
in the extrinsic curvature.
In particular, let us consider first, 
\begin{equation}
S_{(2,a)}
 = \alpha_1 \int_m \sqrt{-\gamma} K^i K_i\,.
\end{equation}

We have
$L_{ab} = 2 \alpha_1 K^i K_{ab\,i}$, and 
$L^{ab}{}_i = 2 \alpha_1 K_i \gamma^{ab}$. Thus
from Eq. (\ref{eq:elec}),
the Euler -Lagrangian derivative is
\begin{equation}
{\cal E}_i =  - 2 \alpha_1 \left[ 
\widetilde{\Delta} K_i +
K^{a b}{}_i K_{a b}{}^j K_j - {1 \over 2} K^j K_j K_i 
\right]\,.
\label{eq:eltec}
\end{equation}
For a quadratic action, the Euler-Lagrange derivative is proportional
to $\tilde\Delta K_i$ plus some cubic in $K_{ab}^i$.
We note that the extremal solutions
$K^i=0$ continue to be solutions of 
this particular theory \cite{CGTZ}.

The linear momentum density takes the form
\begin{equation}
\label{eq:lmomd}
{\cal P}^{a \mu} = \alpha_1 \sqrt{-\gamma}  \left\{
K_i (K^i \gamma^{ab} - 2 K^{ab\,i}) e^\mu{}_{b}
+ 2 \left( \widetilde{\nabla}_a K_i \right) \, n^{\mu\,i}
 \right\}\,.
\end{equation}
Thus,  on all extremal solutions of the
theory defined by
$S = S_{(0)} + S_{(2,a)} $, with 
$K^i=0$, the rigidity makes no contribution to the 
momentum. If the complete action is given by 
the rigidity term, solutions with
$K^i=0$ carry no momentum. 
This  was noted 
by Boisseau and Letelier in \cite{BL} for the special
case of a relativistic string. 

The scale invariance 
of this action in the case of a
relativistic string ($D=2$) can be checked
by verifying that ${\cal P}^{ab} \gamma_{ab} = 0$.

The normal projection of the
linear momentum density is 
\begin{equation}
{\cal P}^{ai} = 2 \alpha_1 \sqrt{-\gamma}
 \widetilde{\nabla}^a K^i\,.
\end{equation}
 Therefore, if the mean curvature is constant, $K^i =$ const.,
so that ${\cal P}^{ai} =0$,
then the equations of motion are of second order
in derivatives of the embedding functions.

The angular momentum density is given by 
\begin{equation}
\label{eq:amtecd}
{\cal M}^{a \mu \nu} = \frac{1}{2}
\; \left[ 
{\cal P}^{a\mu} X^{\nu} - 2 \sqrt{-\gamma} 
K_i n^{\mu \,i} e^{\nu\,a} - 
( \mu \leftrightarrow \nu )\right]\,.
\end{equation}
The second term is proportional to the
bivector normal to $\Sigma$.
Both conditions (\ref{eq:cond1}), (\ref{eq:cond2})
are identically satisfied, so that angular momentum
is conserved.

The total angular momentum 
$M^{\mu\nu}$ also vanishes when $K^i=0$. 
For a DNG string, the well-known Regge inequality
bounds the spin by the mass.
We conclude that for a rigid string, the inequality
 continues to hold
for all DNG solutions \cite{CGTZ}.

We have seen already that for strings the two quadratics are not 
independent so that the $K^i K_i$ theory is, in fact, the 
unique  theory quadratic in extrinsic curvature.
In general, for arbitrary $D$, however we also have
 the other possibility,
\begin{equation}
S_{(2,b)} = \alpha_2 \int_m \sqrt{-\gamma} K_{ab}{}^i K^{ab}{}_i\,.
\end{equation}
 We find easily  that
$L_{ab} = 2 \alpha_2 K^c{}_{a\,i} K_{bc}{}^i$
and $L^{ab}{}_i = 2 \alpha_2 K^{ab}{}_i$. 
From Eq. (\ref{eq:elec}), the Euler-Lagrange
derivative is  
\begin{equation}
{\cal E}_i =
- 2 \alpha_2 \left( \widetilde{\nabla}_a\; \widetilde{\nabla}_b K^{ab}{}_i +
K^{ac}{}_j K^j_{c b} K_{ a}{}^b{}_i 
- {1 \over 2} K_{ab}{}^j K^{ab}{}_j K_i \right)\,. 
\label{eq:1enec}
\end{equation}
This expression for the Euler-Lagrange derivative
is perfectly legitimate.  On the other hand, we
can offer an alternative expression, closer to the
one given for the Euler-Lagrange derivative
of the other quadratic action, Eq. 
(\ref{eq:eltec}).
We use the once contracted Gauss-Codazzi equation,
\begin{equation}
{\cal R}_{ab} - K^i K_{a b\,i} + K_{a c}{}^i K^c{}_{b\,i}=0\,,
\label{eq:gccg}
\end{equation}
and the contracted 
Codazzi-Mainardi integrability condition,
\begin{equation}
\widetilde{\nabla}_a( K^{ a b\,i} 
-\gamma^{a b} K^i ) =  0\,,
\label{eq:codd} 
\end{equation}
to reduce the Euler-Lagrange derivative (\ref{eq:1enec})
 to the form
\begin{equation}
{\cal E}_i =  - 2 \alpha_1 \left( \widetilde{\Delta} K_i +
K^{a b}{}_i K_{a b}{}^j K_j - {1 \over 2} K^j K_j K_i 
- {\cal G}_{ab} K^{ab}{}_i \right) \,,
\label{eq:eltec1}
\end{equation}
where ${\cal G}_{ab} = 
{\cal R}_{ab} - (1/2) {\cal R}\gamma_{ab} $ is the
worldvolume Einstein tensor.
This alternative expression is identical
to Eq. (\ref{eq:eltec}), except for the 
addition of the last term, which vanishes
identically for a string, $D=2$.

The linear momentum density for this theory is
\begin{equation}
{\cal P}^{a\mu } = \alpha_2  \sqrt{-\gamma} 
\left[ \left( K_{cd}{}^i K^{cd}{}_i \gamma^{ab}
- 2 K^a{}_{c\,i} K^{bc\,i} \right) e^\mu{}_b 
+ 2 (\widetilde\nabla_b K^{ab\,i}) n^\mu{}_i \right]\,.
\end{equation}
We can use the Codazzi-Mainardi equation
(\ref{eq:codd}) to simplify the second term, and the
contractions of the Gauss-Codazzi equation (\ref{eq:gcc}),
 (\ref{eq:gccg}), to write the linear momentum
density in the alternative form,
\begin{equation}
{\cal P}^{a\mu } = \alpha_2  \sqrt{-\gamma} 
\left[ \left( K_i K^i \gamma^{ab}
- 2 K_i K^{ab\,i} - 2 {\cal G}^{ab} \right) e^\mu{}_b 
+ 2 (\widetilde\nabla^a K^{i}) n^\mu{}_i \right]\,.
\end{equation}
We note that also for this theory, for 
constant mean curvature solutions, $K^i =$ const.,
the equations of motion are of second order
in derivatives of the embedding functions.
Moreover, for $D=2$, the theory is scale
invariant, since
${\cal G}_{ab} = 0$, identically, so that
 ${\cal P}^{ab} \gamma_{ab} = 0$.

The total angular momentum is
\begin{equation}
M^{\mu\nu} (\Sigma ) = {1 \over 2} \int_{\partial m} 
\eta_a \left[ {\cal P}^{a\mu} X^\nu + 2 \alpha_2 \sqrt{h} K^{ab\,i} 
n^\mu{}_i e^\nu{}_b - 
( \mu \leftrightarrow \nu )\right]\,.
\end{equation}
Whereas in the cases considered previously
the second term, independent of the linear
momentum density, involved only the bivector
normal to $\Sigma$, here it includes also the
bivector normal-tangential to $\Sigma$.

As a last example of an action quadratic in the
extrinsic curvature, we consider now the Einstein-Hilbert
action,
\begin{equation}
\label{eq:ehb}
S_{EH}= \beta \int_m \sqrt{-\gamma} \; {\cal R}\,.
\end{equation}
We have, 
\begin{eqnarray}
L_{ab} &=& 2 \beta \left( K_i K_{ab}{}^i -
K_{ac\,i} K_b{}^{c\,i} \right) = 2 \beta {\cal R}_{ab}\,,
\nonumber \\
L^{ab}{}_i &=& 2 \beta \left( K_i \gamma^{ab}
- K^{ab}{}_i \right)\,.\nonumber
\end{eqnarray}
Note  that the Codazzi-Mainardi equation (\ref{eq:codd}) implies
\[
\widetilde\nabla_b L^{ab}{}_i = 0\,.
\]
This kills the boundary term proportional to $\Phi_i$.

 To calculate the variation of this action, we can use the contracted 
Gauss-Codazzi equation for a flat space-time background, Eq.(\ref{eq:gcc}), 
and the results of the previous section.
We obtain,
\begin{eqnarray}
\delta  S_{EH} 
&=& \beta \int_m \sqrt{-\gamma}\; \left\{ -2 K^{a b\,i}
{\cal G}_{a b} \Phi_i \right\} \nonumber \\
& +&  \beta \int_{\partial m} \sqrt{h} \; \eta_a \left\{ 2 
\left[ K^{a b\,i} - \gamma^{ a b} K^i \right] \widetilde{\nabla}_b \Phi_i  
+ {\cal R} \Phi^a  \right\}\,, 
\end{eqnarray}
where ${\cal G}_{a b}$ is the worldvolume Einstein tensor, 
${\cal G}_{a b}={\cal R}_{a b} -\frac{1}{2} \gamma_{a b} {\cal R}$.
It follows at once that the Euler-Lagrange derivative of
the Einstein-Hilbert action is given by
\begin{equation}
\label{eq:eleh}
{\cal  E}_i =  - 2 \beta K^{a b}{}_i {\cal G}_{a b}\,.
\end{equation}
When $D=2$, for a string, 
the Einstein tensor vanishes identically,
${\cal G}_{ab}=0$ and ${\cal E}_i=0$ identically.
This can also be checked by considering the
difference of the Euler-Lagrange derivatives
for the two actions quadratic in the curvature,
Eqs. (\ref{eq:eltec}), (\ref{eq:eltec1}).

In general, a DNG action with an Einstein-Hilbert correction satisfies the Euler-Lagrange equations,
\begin{equation}
K^{a b}{}_i (\mu \gamma_{ab} + 2\beta 
{\cal G}_{a b})=0\,.
\end{equation}
 In particular, any embedded Einstein manifold, 
satisfying ${\cal G}_{ab}= -(\mu/2\beta) \gamma_{ab}$ 
is a solution.

We note that an Einstein-Hilbert addition to the DNG
action does not change the order of the equations of motion. They
remain second order. 
The action, however, is no longer stationary under the same 
boundary conditions as the DNG theory, even when 
$D=2$. We now need to require that 
$\eta^a \widetilde\nabla_a \Phi^i=0$.
Technically, we trace this necessity to the fact that the action contains a 
term linear in the second derivative of the induced metric.
In general, this spells trouble in a functional integral approach to
the quantum theory\cite{Hawking}. The resolution 
is, of course, well known: 
introduce a surface correction to the action to cancel these
offending surface terms. The 
essence of the variational problem is encountered in the elementary 
description of a free particle by the lagrangian,
$L(x,\ddot x) = - (1/2) x \ddot x$ which is equivalent 
to the lagrangian $ (1/2) \dot x^2$ with the 
addition of a boundary term 
to the corresponding action. This term is simply
$(1/2)  x\dot x$.
To do this in a covariant way for the problem at hand
is a little harder. We are guided by 
the analogous problem in general relativity. 
This issue is treated in Sect. 8.

Here we examine 
the conserved quantitied associated with 
the unadorned Einstein-Hilbert action. With or without surface
modifications, Noether's theorem applies so long as
the action possesses the required invariance.
Noether's theorem is independent of the 
particular boundary conditions that 
are invoked to determine the Euler-Lagrange equations.

The Einstein-Hilbert contribution to the momentum density is given by
\begin{eqnarray}
{\cal P}^{a\,\mu} & = & 
\beta \sqrt{-\gamma} 
\left[ 2 \left( K_{a b}{}^i - \gamma_{ a b} K^i \right) K^{bc}_i 
+ {\cal R}  \right] e^\mu{}_c \nonumber \\
      & = & - 2 \beta \sqrt{-\gamma} 
 {\cal G}^{ab} e^\mu{}_b\,, 
\label{eq:mehb}
\end{eqnarray}
where we have exploited the once contracted 
Gauss-Codazzi relation, 
to obtain the second line. We note that
the vanishing of this momentum density
gives the vacuum Einstein field equations
for the worldvolume. If one is interested
in an embedding formulation of GR in a
flat background spacetime, this observation
isolates the relevant subspace of the 
space of solutions, solving the problem
posed by Regge and Teitelboim in \cite{RT}.
We plan to develop the consequences of this
observation elsewhere. 

The angular momentum density 
${\cal M}^{a\, \mu \nu}$ decomposes 
into a sum of two separately conserved quantities:
\begin{equation}
{\cal M}^{a\,\mu\nu}
= \frac{1}{2} \left[ {\cal P}^{a \mu}  X^{\nu} 
+ 2 \beta \sqrt{-\gamma} \left( K^{ab\,i} 
- \gamma^{a b} K^i \right) n^{\mu}{}_i
e^{\nu}{}_b 
- ( \mu \leftrightarrow \nu ) \right]\,.
\label{eq:momeh}
\end{equation}
Again, both conditions (\ref{eq:cond1}) 
and (\ref{eq:cond2}) are identically satisfied,
so the total angular momentum is conserved.

In the case of a string with worldvolume 
dimension $D=2$, 
the Einstein-Hilbert lagrangian is a topological invariant and
does not contribute to the equations of motion.
We would hope that the corresponding contributions to the 
conserved quantities also vanish.
As expected ${\cal P}^{a\mu} = 0$; however, whereas the first term vanishes in 
${\cal M}^{a\mu \nu}$, the second does not.
However, as we have seen, this non-vanishing part is conserved 
kinematically,  it is divergence free off shell. One is 
always at liberty to add such a tensor to
produce a `new' conservation law. The addition of
the appropriate boundary term, as shown below in Sect. 8,
removes this extra kinematic term.

\section{Linearized equations of motion from 
momenta}

In this section, we consider the second order
variation of the action, and we show how the
linearized equations of motion can be expressed
in terms of the normal  variation of the linear momentum
density ${\cal P}^{a\mu}$. It turns out that this
approach to the linearized equations of motion
is more economical than a direct approach,
which considers the variation of the Euler-Lagrange derivative, $\delta {\cal E}^i (L) $ 
\cite{GV,Guven,CarterDNG,FL,Defo}. The virtue
of this approach is that it involves the
variation of geometrical quantities of 
lower order in derivatives of the
embedding functions.

In Sect. 3, we have shown explicitly how
to express the equations of motion in terms
of the conservation of the appropriate
projections of the momenta. 
The key equation we used, for an infinitesimal
translation, is Eq. (\ref{eq:gfdat1}). We 
consider only its normal variation for the
moment. We obtain
\begin{equation}
\delta_\perp ( \delta S ) =
\epsilon_\mu \left\{ 
\int_m \sqrt{-\gamma} \left[ K^j \Phi_j
 {\cal E}^i n^\mu{}_i + ( \tilde\delta_\perp {\cal E}^i )
 n^\mu{}_i -  {\cal E}^i (\widetilde\nabla_a \Phi_i )
e^{\mu \, a} \right] 
+ \int_m \nabla_a \delta_ \perp {\cal P}^{a\mu}\right\} \,.
\label{eq:delta2}
\end{equation}
Here we have used Eq. (\ref{eq:sqk}), and the
second normal deformation Gauss-Weingarten 
equation \cite{Defo},
\begin{equation} 
\tilde\delta_\perp n^\mu{}_i =
( \widetilde\nabla_a \Phi^i ) e^{\mu\,a}\,.
\end{equation}
In the last term,
we have
exploited the fact that, since ${\cal P}^{a\mu}$ is
a vector density of weight one, variation and 
covariant differentiation commute,
$\delta \nabla_a {\cal P}^{a\mu}
= \nabla_a \delta {\cal P}^{a\mu}$.

When the action is invariant under translations,
$\delta S = 0$, this equation relates the normal
variation of the Euler-Lagrange derivative to the
divergence of the normal variation of the linear
momentum density. 

Following the same
strategy we used for the
equations of motion, let us decompose 
${\cal P}^{a\mu}$ into its 
worldvolume projections,
\begin{equation}
\delta_\perp {\cal P}^{a\mu} = [ \delta_\perp
{\cal P}]^{ab} e^\mu{}_b
+ [ \delta_\perp {\cal P}]^{a i} n^\mu{}_i\,,
\end{equation}
where we use the brackets notation in order to distinguish
the projection of the variation from the variation of
the projection, since in general they are different,
{\it e.g.} $[\delta_\perp
{\cal P}]^{ab} \neq \delta_\perp
{\cal P}^{ab}$. In fact, this is the reason that
makes the approach described here more efficient
than a direct approach.

A straightforward calculation gives that when the action
is invariant under translations, so that $\delta S = 0$
on the left hand side of Eq. (\ref{eq:delta2}), 
the worldvolume projections of Eq. (\ref{eq:delta2})
are
\begin{eqnarray}
\sqrt{-\gamma} ( \tilde\delta_\perp {\cal E}^i ) &=&
- \sqrt{-\gamma} K^j \Phi_j {\cal E}^i 
- \widetilde\nabla_a  [ \delta_\perp {\cal P}]^{a i}
+ [ \delta_\perp {\cal P}]^{ab} K_{ab}{}^i \,,
\label{eq:lin1}
\\
\sqrt{-\gamma} {\cal E}^i (\widetilde\nabla^b \Phi_i )
&=&   \widetilde\nabla_a  [ \delta_\perp {\cal P}]^{a b}
+ [ \delta_\perp {\cal P}]^{ai} K_a{}^b{}_i \,.
\label{eq:lin2}
\end{eqnarray} 
This latter equation is merely an identity, as 
expected from reparametrization invariance.
The first equation can be used to express
the linearized equations of motion, about a
solution of the equations of motion, in the form
\begin{equation}
\widetilde\nabla_a  [ \delta_\perp {\cal P}]^{a i}
- [ \delta_\perp {\cal P}]^{ab} K_{ab}{}^i 
= 0\,.
\end{equation}
Note the apparent similarity with the equations of motion
expressed in the form (\ref{eq:eom3}). 
The usefulness of this expression is that in general,
the variation of the momentum density is easier to calculate
than the variation of the Euler-Lagrange derivative.

The variation of the total linear momentum is then
simply
\begin{equation}
\delta_\perp P^\mu (\Sigma ) = 
\int_{\partial m} \eta_a  \delta_\perp {\cal P}^{a\mu}\,.
\end{equation}
Again, it should be emphasized that this
 is much easier than a direct variation
of $P^\mu (\Sigma )$.

In order to illustrate this  approach, 
and  to
give a simple application, let us consider a DNG
object. 
The linear momentum density is given by Eq. (\ref{eq:lddng}). We can use Eqs. (\ref{eq:dgab1}), 
(\ref{eq:sqk}),
together with
the first Gauss-Weingarten deformation equation
(see Ref. \cite{Defo})
\begin{equation}
\delta_\perp e^\mu{}_a = K_{ab}{}^i \Phi_i
e^{\mu\,b} + (\widetilde\nabla_a \Phi^i ) n^\mu{}_i
\,,
\end{equation}
 so that
\begin{equation}
\delta_\perp {\cal P}^{a\mu} = - \mu \sqrt{-\gamma} 
[ ( K_i \gamma^{ab} - K^{ab}{}_i ) \Phi^i 
e^\mu{}_b + ( \widetilde\nabla^a \Phi^i ) n^\mu{}_i ]\,.
\end{equation}
We read off the projections,
\begin{equation} 
[ \delta_{\perp} {\cal P} ]^{ab} = - \mu \sqrt{-\gamma}
( K_i \gamma^{ab} - K^{ab}{}_i ) \Phi^i \,,
\end{equation} 
\begin{equation}
[ \delta_{\perp} {\cal P} ]^{ai} = - \mu \sqrt{-\gamma} 
\widetilde\nabla^a \Phi^i \,. 
\end{equation}

Recalling that, for a DNG object, ${\cal E}_i = - \mu
K_i$, and substituting in Eq. (\ref{eq:lin1}), we find
\begin{eqnarray}
( \tilde\delta_\perp K^i ) &=&
- K^j \Phi_j K^i 
- \widetilde\nabla_a  ( \widetilde\nabla^a \Phi^i ) 
+ ( K_j \gamma^{ab} - K^{ab}{}_j ) \Phi^j K_{ab}{}^i \nonumber \\
&=& - \widetilde\nabla_a  ( \widetilde\nabla^a \Phi^i ) 
-  K_{ab}{}^i K^{ab}{}_j  \Phi^j\,,
\end{eqnarray}
which agrees with the expression derived {\it
e.g.} in Ref. \cite{Defo} (see also 
\cite{Guven,CarterDNG,FL}).

Let us now confirm
that Eq. (\ref{eq:lin2}) is in fact an 
identity. Substitution of the projections gives
\[
 K^i (\widetilde\nabla^b \Phi_i )
 =   \widetilde\nabla_a  [ ( K_i \gamma^{ab} - K^{ab}{}_i ) \Phi^i ]
+  ( \widetilde\nabla^a \Phi^i ) K_a{}^b{}_i \,,
\]
and this expression is identically zero, as follows
from the Codazzi-Mainardi integrability condition,
Eq. (\ref{eq:codd}).

This example renders transparent the advantages of this
approach. Rather than dealing with the variation of
the extrinsic curvature, all one needs here is the
variation of the intrinsic geometry of the worldvolume.

What about the second variation 
parallel to the worldvolume,
$\delta_\parallel ( \delta S )$? As expected from 
reparametrization invariance, it adds nothing new.
We show this explicitly in Appendix B for a DNG
brane.

\section{Einstein-Hilbert action with
surface term}

In this section, we consider the addition 
of a surface term to the Einstein-Hilbert
brane action we have briefly described
at the end of Sect. 6, 
\begin{equation}
\label{eq:eh1}
S = S_{EH} + S_{\partial m} \,,
\end{equation}
where the first term is defined in Eq. (\ref{eq:ehb}), and
\begin{equation}
S_{\partial m} =
2 \beta \int_{\partial m} \sqrt{h}  \,\kappa \,,
\label{eq:pm}
\end{equation}
and $\kappa$ is the trace of the 
extrinsic curvature $\kappa_{AB}$ of the spacelike boundary provided by
$\Sigma_{(i)}$ and $\Sigma_{(f)}$ embedded in $m$. 

An analogous term is added to the 
Einstein-Hilbert action in general relativity in order that
the variational principle applied to the action 
yield the Einstein equations in a bounded region 
subject to the boundary condition that the 
metric induced on the boundary is fixed and no more \cite{York,Hawking}. Technically, 
the variation of the boundary term precisely cancels normal 
derivatives of the variation of the metric tensor on the boundary,
which occur in the variation of the Einstein Hilbert action.
In general relativity this term is diffeomorphism invariant. Here, 
it is also Poincar\'e invariant.  
Interestingly enough, the particle analogue mentioned 
in Sect. 6
is incomplete in this respect. Whereas the 
original lagrangian $L_0(x,\dot x) = (1/2) \dot x^2 $ is
invariant under translations, the lagrangian $L(x,\ddot x) =
-  (1/2) x \ddot x$ is not and so 
Noether's theorem cannot be applied 
directly to it.  The boundary term is necessary to restore 
the translation invariance of the problem.

The variation of Eq.(\ref{eq:pm}) poses new 
technical difficulties. In general relativity, the 
dynamical variables to be varied are the 
spacetime metric coefficients. In the present context, 
we need to vary the embedding functions describing the worldvolume. 
Fortunately, the relevant formalism
has been developed in Refs. \cite{edges,Defoedges},
for the case of a timelike boundary. Its adaptation
to the case of interest here requires only some
minor sign modifications.

The problem is simplified by treating the
 boundary of the 
worldvolume  as two embedded 
spacelike surfaces in the background 
Minkowski space-time described by Eq. (\ref{eq:end3}).
We denote the normals to $\partial m$ in spacetime by
$m^\mu{}_I$ ($I,J, \cdots = 0,1,2,\cdots,N-D$). It
is also convenient to exploit a
normal basis which is adapted
to the worldvolume $m$: we choose the basis
$m^{\mu\,I} = \{ \eta^{\mu}, n^{\mu}{}_i \}$, 
supplementing
the normals $n^{\mu}{}_i$ to $m$ in spacetime 
with  $\eta^{\mu} = e^\mu{}_a \eta^a$,
 the normal to the boundary which is tangent to
the worldvolume.
Let $L_{AB}{}^I$ represent the extrinsic curvatures associated with the
embedding $ X$. With respect to the adapted
basis, 
$L_{AB}{}^i = K_{ab}{}^i \epsilon^a{}_A \epsilon^b{}_B =
K_{AB}{}^i$, 
 and
$L_{AB}{}^0 = \kappa_{AB}$.
(For details see \cite{edges,Defoedges}.

Let us examine an arbitrary deformation of
$\partial m$ in spacetime. 
If $\partial m$ is closed (as we assume), we need only 
consider a normal deformation of $\partial m$. 
Let us first expand 
\begin{equation}
\delta_{\bar\perp} X^\mu = \Phi^I  m^\mu{}_{I}
= \psi \eta^\mu + \phi^i n^\mu{}_i \,. 
\end{equation}
We use the symbol $\bar\perp$ to distinguish
this normal  variation, which includes a variation
along $\eta^\mu$, from the worldvolume normal variation 
used earlier. We have also expressed the variation
with respect to 
the adapted basis, so that $\Psi^0 = \psi,
\Psi^i = \psi^i $.
 
We now exploit the formalism developed in 
Ref.\cite{Defoedges} to express the induced normal variations
of $h_{AB}$ and $L_{AB}{}^I$ as follows:
\begin{eqnarray}
\delta_{\bar\perp}  h_{AB} &=& 
2 L_{A B}{}^I 
\Phi_I  \,, \\
\delta_{\bar\perp} L_{AB}{}^I  
&=& -\hat{\cal D}_A \hat{\cal D}_B \Phi^I + 
L_{A}{}^{C\,I}  L_{ CB\,J} 
\; \Phi^J
+ \hat{\gamma}^I_J L_{A B}{}^J \,.
\label{eq:labi} 
\end{eqnarray} 
Here ${\cal D}_A$ is the covariant 
derivative compatible with $h_{AB}$, 
and $\hat{\cal D}_A$ its extension that is 
covariant under rotations of normals to $\partial m$ in spacetime.
We need to compute the variation for the component $I=0$ 
in this last expression. 
The boundary term is not covariant 
under rotations of the normals, $m^\mu{}_I$: $\kappa_{AB}$ is the 
extrinsic curvature which corresponds to $\eta^\mu$,
 there is 
no rotational arbitrariness here. 
For this reason, the appropriate deformation operator 
we apply to $\kappa$ is $\delta_{\bar\perp} $ and
not the manifestly rotationally 
covariant 
object $\tilde\delta_{\bar\perp}$ \cite{edges}. 
This implies the necessity to restore the 
deformation connection
$\hat{\gamma}^I{}_J$ on the right hand side of Eq.(\ref{eq:labi}). 
This would spell disaster if we needed to evaluate all
of $\hat\gamma_{IJ}$. However, we only require 
$\hat \gamma_{0i}$ which {\it is} well defined on $\partial m$.
We then find \cite{Defoedges}
 \begin{eqnarray}
\delta_{\bar{\perp}}  \kappa &=& -{\cal D}_A {\cal D}^A \psi 
+ \left( K^{Ai} K_{Ai} - \kappa^{A B} \kappa_{A B} 
-    h^{A B} K_{A B}{}^i  K_{a b\,i} 
\eta^a \eta^b \right) \psi  \nonumber \\  
&-& {\cal D}_A \left( K^{i A} \phi_i \right) +
\left(   \eta^a h^{ A B}K_{A B }{}^i 
- \epsilon^{a A} K_{A}{}^i  \right)  
\tilde{\nabla}_a \phi_i 
- \kappa^{AB} K_{AB}{}^i \phi_i \,.
\label{eq:dk3}
\end{eqnarray} 
To arrive at this expression we have used the 
fact that $\hat{\gamma}^{0i}= - 
\eta^a \eta^b K^i_{a b}\psi
+\eta^a \widetilde{\nabla}_a \phi^i$ and 
we have introduced the notation, 
$ K_A{}^i = \eta^a \epsilon^b{}_A K^i_{ab}$.
Some simplification is possible.
Let us begin with the terms linear in  $\psi$.
We can use the contracted Gauss-Codazzi equation
to obtain,  
\[
K^{Ai} K_{Ai} - h^{A B} K_{ A B }{}^i K_{a b}{}_i 
\eta^a \eta^b
=  - \eta^a \eta^b {\cal R}_{ab}\,.
\]
In addition, we have
\[
\left(   \eta^a h^{ A B}K_{A B }{}^i 
- \epsilon^{a A} K_{A}{}^i  \right)  
\widetilde{\nabla}_a \phi_i 
= \eta_a
\left( K^i \gamma^{ab}- K^{ab\,i} \right) \widetilde{\nabla}_b \phi_i \,.  
\]
Inserting these partial results into Eq. (\ref{eq:dk3}),
we obtain
\begin{eqnarray}
\delta_{\bar{\perp}} \kappa &=& 
-{\cal D}_A {\cal D}^A \psi
- \left( \kappa^{A B} \kappa_{A B} + \eta^a \eta^b 
{\cal R}_{ab} \right) \psi \nonumber \\
&-& {\cal D}_A \left( K^{i A} \phi_i \right)
 + \eta_a
\left( K^i \gamma^{ab}- K^{ab\,i} \right) 
\widetilde{\nabla}_b \phi_i - \kappa^{AB} K_{A B}{}^i \phi_i \,.
\end{eqnarray} 

We have now all the ingredients needed for the 
calculation of the normal variation of the
boundary action (\ref{eq:pm}). We neglect
total derivatives, and we find
\begin{eqnarray}
\delta_{\bar{\perp}} S_{\partial m} &=&
 2 \beta \int_{\partial m} \sqrt{h} 
 \{  \left[ \kappa^2 - \kappa_{AB} \kappa^{AB}
- \eta^a \eta^b {\cal R}_{a b} \right] \psi 
\nonumber \\
 &+&  ( h^{AB} \kappa - \kappa^{AB} ) K_{ AB}{}^i \phi_i 
+ \eta_a 
\left( K^{i} \gamma^{ab} - K^{ab\,i} \right) \widetilde{\nabla}_b \phi_i
\}\,.
\end{eqnarray}
To establish contact with the bulk variation, 
we must identify $\psi= \eta^a\Phi_a$.
The total boundary contribution, obtained by 
summing this expression with the second line 
of Eq. (\ref{eq:mehb}), therefore does not
contain any term involving derivatives
of the $\Phi^i$.

Let us consider now a translation of the boundary in the background 
space-time. 
The linear momentum associated with this contribution is then 
given by
\begin{equation}
\label{eq:momsur}
p^{\mu} (\Sigma ) = 2 \beta \int_\Sigma \sqrt{h} 
\;  \left\{ \eta^\mu  \left[ \kappa_{A B} 
\kappa^{A B} - \kappa^2
+ \eta^a \eta^b {\cal R}_{a b} \right]
+ \eta^a {\cal R}_{ab} e^{\mu\,b} 
+  K_{AB}^i  ( h^{AB} \kappa - \kappa^{AB} )  n^{\mu}{}_i \right\}\,.
\end{equation}
It should be noticed that the
surface momentum has an additional normal component.
The linear momentum for the total action (\ref{eq:eh1})
using Eq. (\ref{eq:mehb}), becomes
\begin{equation}
P^\mu (\Sigma ) + p^\mu (\Sigma ) = \beta
\int_\Sigma \sqrt{h}
 \left\{ \eta^\mu  \left[ \kappa_{AB} 
\kappa^{A B} - \kappa^2
+ \eta^a \eta^b {\cal G}_{a b} \right]
+  K_{AB}^i  (h^{AB} \kappa - \kappa^{A B} )  n^{\mu}{}_i \right\} \,.
\label{eq:totp}
\end{equation}
Note that this expression vanishes identically
for a string, since for $D=2$, we have ${\cal G}_{ab} = 0$,
and moreover in the degenerate case of
a one dimensional boundary, $k_{AB} = h_{AB} k$.

Consider now an infinitesimal Lorentz transformation.
The boundary contribution to the
angular momentum is given by 
\begin{equation}
m^{\mu \nu} (\Sigma ) = 
{1 \over 2} \int_\Sigma  \left\{ 
\tilde{p}^{\mu} X^{\nu} + 2 \beta 
\sqrt{h} \eta_a \left( K^{i} \gamma^{ab} - K^{ab\,i} \right) n^\mu{}_i e^\nu{}_b
 -\left( \mu \leftrightarrow \nu \right) 
 \right\} \,,
\end{equation}
where the quantity $\tilde{p}^{\mu}$  in the
first term is the integrand appearing in Eq.(\ref{eq:momsur}).

The second term is what is needed to cancel the
offending term in the angular momentum for the
bulk Einstein-Hilbert action, in Eq. (\ref{eq:momeh}), 
so that
the angular momentum for the total action       
is now
\begin{equation}
M^{\mu \nu} (\Sigma ) + m^{\mu\nu} (\Sigma)
= {1 \over 2} \int_\Sigma  \left[ \Pi^\mu  X^{\nu} - ( \mu \leftrightarrow \nu )
\right]\,,
\end{equation}
where we denote with $\Pi^\mu$ the integrand of the
total linear momentum (\ref{eq:totp}). 

In conclusion, the same surface term that is
appropriate to lower the order of the boundary
conditions, also cancels the kinematic term in the
angular momentum density.

\section{Discussion}

In this paper, we have presented a new 
approach to the derivation of the linear
and angular momentum  for a brane propagating
in Minkowski spacetime based on the worldvolume geometry. 
We have considered a large class of brane actions, depending
on the intrinsic and as well on the 
extrinsic geometry of the worldvolume,
up to a first derivative of the extrinsic curvature.
The generalization to a more general action is
straightforward, following the guidelines 
given in the paper. 

This analysis may be extended in  a
straightforward way to treat the 
corresponding conserved quantities for a brane 
propagating on a background spacetime possessing 
Killing vector fields.

A particular simple case we have not discussed explicitly here
is the degenerate case of a point object described by a
higher order action. We will discuss this case elsewhere.

It would also be interesting to
apply this geometrical approach
to supersymmetric branes \cite{superbrane}.
While the geometry of such objects is well understood, 
to our knowledge, the geometry of deformations of 
superembedded surfaces remains to be developed. 

\vskip 1cm

\noindent{\bf Acknowledgements}
\vskip .5cm
We have received partial support 
from CONACyT grant no. 211085-5-0118PE.
Our thanks to Charles Torre for useful suggestions.
\newpage

\noindent{\bf Appendix A}

\vskip .5cm

In this section, we consider the more general case
of an action that depends also on first derivatives
of the extrinsic curvature, 
\begin{equation}
S_{(ho)} [Y] = \int_m \sqrt{-\gamma} L ( \gamma^{ab},
K_{ab}{}^i , \widetilde\nabla_a K_{bc}{}^i )\,.
\end{equation}
We can recycle the results of Sect. 5, for the
variation with respect to the first arguments,
that will not be repeated here. 
The new part is
\begin{equation}
\delta_\perp S_{(ho)} = \int_m L^{abc}{}_i \tilde\delta_\perp
\widetilde\nabla_a K_{bc}{}^i \,,
\label{eq:newprp}
\end{equation}
where we have defined    
\begin{equation}
L^{abc}{}_i = {\partial L \over \partial \widetilde\nabla_a K_{bc}{}^i}
\,.
\end{equation}
Note that, as a consequence of the Codazzi-Mainardi
integrability condition, and the symmetry of $K_{ab}{}^i$,
we have the symmetry property
\begin{equation}
\widetilde\nabla_a K_{bc}{}^i = \widetilde\nabla_{(a} K_{bc)}{}^i\,,
\end{equation}
from which it follows that
\begin{equation}
L^{abc}{}_i = L^{(abc)}{}_i\,.
\end{equation}

In order to evaluate Eq. (\ref{eq:newprp}), we need
to commute $\tilde\delta_\perp$ with $\tilde\nabla_a$.
To do this, we need the following expressions (see also \cite{Defo})
\begin{eqnarray}
\delta_\perp \gamma_{ab}{}^c &=&
\gamma^{cd} \left[ \nabla_a 
\left( K_{bd}{}^i \Phi_i \right)
+ \nabla_b 
\left( K_{ad}{}^i \Phi_i \right)
- \nabla_d 
\left( K_{ab}{}^i \Phi_i \right)
\right]\,,
\\
\delta_\perp \omega_a{}^{ij} 
&=& K_a{}^{b\,i} \widetilde\nabla_b \Phi^j
- K_a{}^{b\,j} \widetilde\nabla_b \Phi^i\,.
\end{eqnarray} 
Now, it is a direct computation to
obtain that the contribution of (\ref{eq:newprp})
to the boundary term of the normal 
variation of the action is given by,
\begin{eqnarray}
\delta_\perp S_{(ho)} &=&
\int_{\partial m} \sqrt{h} \eta_a
[ - \left( \widetilde\nabla_b \widetilde\nabla_c
L^{abc}{}_i \right) \Phi^i
+ \left( \widetilde\nabla_b L^{abc}{}_i \right)
\widetilde\nabla_c \Phi^i 
- L^{abc}{}_i \widetilde\nabla_b \widetilde\nabla_c \Phi^i
\nonumber \\
&-& L^{bcd}{}_i K_{bc}{}^j K^a{}_{d\,j} \Phi^i 
+ 3 L^{bcd}{}_j K_{bc\,i} K_d{}^{a\,j} \Phi^i
- 3 L^{abc}{}_j K_b{}^{d\,j} K_{cd\,i} \Phi^i
] \,.
\end{eqnarray}
If this seems complicated, well the Euler-Lagrange
derivative is worse.

For an infinitesimal translation, we obtain
that the total contribution to the momentum density
is
\begin{eqnarray}
{\cal P}^{a\mu} &=& \sqrt{-\gamma} 
\{ 
\left[ L_{(ho)} \delta^{a}{}_d 
- L^{abc}{}_i (\widetilde\nabla_b K_{cd}{}^i ) 
+ (\widetilde\nabla_b L^{abc}{}_i )
K_{cd}{}^i \right] e^{\mu\,d}
\nonumber \\
&+& [
-  \left( \widetilde\nabla_b \widetilde\nabla_c
L^{abc}{}_i \right) 
- L^{bcd}{}_i K_{bc}{}^j K^a{}_{d\,j}  
+ 3 L^{bcd}{}_j K_{bc\,i} K_d{}^{a\,j} 
\nonumber \\
&-& 2 L^{abc}{}_j K_b{}^{d\,j} K_{cd\,i} 
] n^{\mu\,i} 
\}
\,.
\end{eqnarray}

The contribution to the angular momentum density 
is
\begin{eqnarray}
{\cal M}^{a\mu\nu} =
{1\over 2} \{ {\cal P}^{a\mu} X^\nu
&+& \sqrt{-\gamma} [
(\widetilde\nabla_b L^{abc}{}_i ) n^{\mu\,i}
e^\nu{}_c  - 2 L^{abc}{}_i K_{cd}{}^i  e^{\mu\,d}
e^\nu{}_b \nonumber \\
&+& L^{abc}{}_i K_{bc}{}^j n^{\mu\,i} n^{\nu}{}_j ]
- (\mu \leftrightarrow \nu ) \} \,.
\nonumber
\end{eqnarray}
We note that in this expression all of the bivectors
enter.

\vskip 1cm

\noindent{\bf Appendix B}

\vskip .5cm

In this appendix, we show explicitly
that, as stated at the end of Sect. 7,
the vanishing of 
the second variation parallel to the worldsheet,
$\delta_\parallel (\delta S )$ results in mere 
identities.
 
The parallel variation of Eq. (\ref{eq:gfdat1}) 
gives
\begin{equation}
\delta_\parallel ( \delta S ) =
\epsilon_\mu 
\int_m \nabla_a \left[\sqrt{-\gamma} \Phi^a
 {\cal E}^i n^\mu{}_i 
+ \delta_ \parallel {\cal P}^{a\mu}\right] \,,
\label{eq:deltapar}
\end{equation}
where we have used Eq. (\ref{eq:tpda}), valid for any
worldvolume scalar density of weight one.
We decompose $\delta_\parallel
{\cal P}^{a\mu}$ in terms of its
worldvolume projections,
\[
\delta_\parallel {\cal P}^{a\mu} = [ \delta_\parallel
{\cal P}]^{ab} e^\mu{}_b
+ [ \delta_\parallel {\cal P}]^{a i} n^\mu{}_i\,.
\]
Substituting into Eq. (\ref{eq:deltapar}), and using
the Gauss-Weingarten equations (\ref{eq:gw1}),(\ref{eq:gw2}),
we find the parallel analogue of Eqs. (\ref{eq:lin1}),
(\ref{eq:lin2}),
\begin{equation}
\nabla_a \left( \sqrt{-\gamma}\; 
\Phi^a {\cal E}^i \right) = 
- \widetilde\nabla_a  [ \delta_\parallel {\cal P}]^{a i}
+ [ \delta_\parallel {\cal P}]^{ab} K_{ab}{}^i \,,
\label{eq:plin1}
\end{equation}
\begin{equation}
\sqrt{-\gamma} \Phi^a {\cal E}^i K_a{}^b{}_i 
=  - \widetilde\nabla_a  [ \delta_\parallel {\cal P}]^{a b}
- [ \delta_\parallel {\cal P}]^{ai} K_a{}^b{}_i \,.
\label{eq:plin2}
\end{equation} 
Both these equations are mere identities. Let us confirm
this for the special case of a DNG object.
Using the parallel deformation Gauss-Weingarten equation
(see \cite{Algebra}),
\[
\delta_\parallel e^\mu{}_a = (\nabla_a \Phi^b ) e^\mu{}_b
+ K_{ab}{}^i \Phi^b n^\mu{}_i\,,
\]
we have that
the parallel variation of Eq. (\ref{eq:lddng})
gives 
\[
\delta_\parallel {\cal P}^{a\mu}
= - \mu \sqrt{-\gamma} \left[ \left( \nabla_c \Phi^c
\gamma^{ab} - \nabla^b \Phi^a \right) e^\mu{}_b
- K^{ab\,i} \Phi_b n^\mu{}_i \right]\,,
\]
so that the worldvolume projections are
\[
[ \delta_{\parallel} {\cal P} ]^{ab} = 
- \mu \sqrt{-\gamma}  \left( \nabla_c \Phi^c
\gamma^{ab} - \nabla^b \Phi^a \right)\,,
\]
\[
[ \delta_{\parallel} {\cal P} ]^{ai} = 
\mu \sqrt{-\gamma} K^{ab\,i} \Phi_b \,. 
\]
Substituting into Eq. (\ref{eq:plin1}) gives
\[
\nabla_b ( \Phi^b K^i ) = \nabla_a [ K^{ab\,i} \Phi_b ]
+ ( \nabla_c \Phi^c
\gamma^{ab} - \nabla^b \Phi^a )
K_{ab}{}^i\,. 
\]
Using the contracted Codazzi-Mainardi integrability
condition, Eq. (\ref{eq:codd}), one can easily verify that
this equation is satisfied identically.
On the other hand, substitution of the projections
into Eq. (\ref{eq:plin2}) gives
\[
\Phi^a K_i K_a{}^{b\,i} =
\Phi_c K^{ac\,i} K_a{}^b{}_i 
- \nabla_a \left(  \nabla_c \Phi^c
\gamma^{ab} - \nabla^b \Phi^a \right)\,.
\]
This equation can be seen to vanish identically as
well. This requires the contracted Gauss-Codazzi
equation (\ref{eq:gcc}), for the lefthand side
together with the first term on the right hand side,
and the Ricci identity for the remaining two terms,
\[
\left(\nabla_a \nabla_b - \nabla_b \nabla_a 
\right) \Phi^b = - {\cal R}_{ab} \Phi^b\,.
\]

\end{document}